\documentclass[aps,prd,reprint,superscriptaddress]{revtex4-2}
\usepackage{xcolor}
\usepackage{graphicx}
\usepackage{dcolumn}
\usepackage{bm}
\usepackage{hyperref}
\usepackage{graphicx, amsmath, amssymb, amstext,xcolor,txfonts}
\usepackage{relsize, float, multirow}
\usepackage{array, epstopdf, hyperref}
\usepackage{caption, subcaption, mathtools}
\usepackage{url}
\usepackage{newtxmath,ulem}

\definecolor{seagreen}{rgb}{0.18, 0.55, 0.34}

\begin{document}


\title{Minkowski Functionals of [C{\sc ii}] Line Intensity Maps: Probing Astrophysics and Primordial Features}



\author{Sandeep Kumar Acharya}
\email{sandeepkumaracharya92@gmail.com}

\author{Pravabati Chingangbam}%
\affiliation{Indian Institute of Astrophysics, Koramangala II Block, Bengaluru 560 034, India
}%

\date{\today}

\begin{abstract}
Line intensity mapping aims to measure the total cumulative light of spectral lines from galaxies,  without resolving individual sources. In this work, we investigate mock intensity maps of ionized carbon ([C\,{\sc ii}]) using Minkowski functionals, which capture the morphological properties of the maps. We generate semi-numerical simulations of [C\,{\sc ii}] intensity maps at $z\approx 6$ and 7.6, adopting an astrophysical model that relates [C\,{\sc ii}] line luminosity to the mass of the host dark matter halos. We find that Minkowski functionals can provide tight constraints on the astrophysical model parameters in the presence of noise characteristics expected from the upcoming {\it CCAT} survey. We further investigate the prospects of probing a bump-like feature superimposed on the standard single power-law primordial power spectrum. Using a Fisher matrix formalism, we demonstrate that the effects of this primordial feature can be disentangled from those of the astrophysical model parameters. Combining measurements across multiple smoothing scales as well as redshifts can further improve the constraints and enhance the prospects for detecting such primordial features.
\end{abstract}

\maketitle


\section{Introduction}
The remarkable precision of Cosmic Microwave Background (CMB) measurements \citep{WMAP2011,Planck2020,ACT2025,SPT2025} and large-scale galaxy surveys \citep{eBOSS2021,DESI2025} has enabled percent-level constraints on the standard $\Lambda$CDM cosmological model. Nevertheless, several fundamental questions such as the nature of dark matter, the initial conditions of the Universe and the physics of inflation etc., remain open. Addressing these questions requires a detailed understanding of the formation and evolution of cosmic structures, including galaxies, across cosmic time.

CMB observations provide an essentially two-dimensional, redshift integrated information  of the Universe along the line of sight. In contrast, spectroscopic galaxy surveys provide three-dimensional maps of discrete galaxies by measuring both their angular positions and redshifts. With increasingly large galaxy surveys, this enables the evolution of the large-scale structure to be mapped across cosmic time. However, obtaining spectroscopic redshifts for large numbers of galaxies becomes increasingly challenging at higher redshifts, where galaxies are intrinsically fainter. Accurate spectroscopic redshift measurements require sufficiently detailed spectra with many frequency channels, making such observations expensive and difficult to scale to the enormous galaxy samples required by next-generation surveys.

Line intensity mapping (LIM) offers a complementary approach to tracing the evolution of cosmic structures without requiring the detection and spectroscopic identification of individual galaxies. Instead of resolving individual sources, LIM measures the aggregate emission from large populations of galaxies in a given spectral line, producing a three-dimensional intensity field with angular and redshift information. By targeting different atomic and molecular emission lines, such as CO, [C\,{\sc ii}], 21 cm and Ly$\alpha$, LIM can probe a wide range of redshifts and physical processes associated with galaxy formation and evolution. We refer the interested reader to \cite{BK2022,CL2026} for extensive discussions on this topic.  While CO and [C\,{\sc ii}] emission are expected to track star formation inside galaxies, 21cm and Ly$\alpha$ lines are more sensitive to intergalactic medium properties such as its temperature and neutral hydrogen abundance. This makes LIM a promising technique for efficiently mapping large cosmological volumes and studying the growth of large-scale structure across cosmic time, while potentially accessing epochs and scales that are difficult to probe with conventional galaxy surveys. Some proposed and ongoing experiments or collaboration include: COPSS \citep{COPSS2015}, COMAP \citep{COMAP2022} for CO, CCAT \citep{CCAT}, CONCERTO \citep{CONCERTO2020}, TIME \citep{TIME2014} for [C\,{\sc ii}], Meerkat \citep{Meerkat2023}, CHIME \citep{CHIME2025} for 21~cm, HETDEX \citep{HETDEX2008} for Ly$\alpha$, SPHEREX \citep{SPHEREX2014} for OII, OIII, Ly$\alpha$ line.

In the $\Lambda$CDM model, galaxies form in the gravitational potential well of dark matter halos. The spatial distribution of dark matter halos capture the particle property of dark matter as well as the initial condition of the universe that was set during inflationary epoch. The spatial fluctuations of 
LIM maps are correlated with the distribution of galaxies, which in turn act as a proxy for dark matter halos. Therefore, LIM maps capture cosmological information as well as intrinsic astrophysical information regarding specific line emissions from galaxies. 
The prospects for constraining the properties of dark matter \citep{Kovetz2019,BMHPL2021,BCVK2021}, primordial non-Gaussianity from the inflationary epoch \citep{CBCHM2025}, the expansion history of the Universe \citep{BBK2019}, and the epoch of reionization \citep{CVCW2020,MMD2021,RPBS2025} using LIM maps have been extensively studied. To extract robust cosmological information from LIM observations, it is essential to account for and, where possible, break degeneracies between cosmological and astrophysical parameters.

The power spectrum, or equivalently the two-point correlation function, is the most commonly used statistic for characterizing the clustering properties of LIM maps and their dependence on  cosmological and astrophysical parameters \citep{CPPM2008,VL2010,VTL2011,GCSSL2011,Carilli2011,BK2022,CL2026}. However, LIM maps are expected to be non-Gaussian, owing to the nonlinear clustering of matter and the discrete distribution of galaxies and dark matter halos, as well as the complex relationship between halo properties and line luminosity. 
The authors of \cite{BKBDK2017,SBSK2024} studied the one point probability distribution or voxel intensity distribution of LIM maps and demonstrated that combining this statistic with the power spectrum 
can provide tighter constraints on cosmological and astrophysical parameters. This highlights the potential of higher-order and non-Gaussian statistics to extract additional information from LIM observations that is not captured by the power spectrum alone.
With this motivation,  in this work we study the LIM maps using Minkowski functionals (MFs) which are geometrical and topological statistics that quantify the shape and connectivity of a smooth random field. These statistics have been applied to a variety of cosmological data, ranging from the CMB (see e.g.\cite{GPJBBBS1990,SG1998,Hikage:2012,MSCRHC2013,Ducout:2013,Ganesan:2015,Santos:2016,Ade:2015ava,Chingangbam:2017sap,Carones:2024,Bashir:2026}),  Galactic foregrounds \cite{Chingangbam:2013,Rana:2018oft,RCG2021,Rahman:2022edq,Martire:2023ytg,Ranucci:2026},  large scale structure data \citep{MLZHN2019,APHHK2020,LYYZ2020,LJF2022,SHH2025,GACP2026},  21 cm simulations from the epoch of reionization  \citep{Gleser:2006,Yoshiura:2016,Chen:2019,VNC2026}, to  interGalactic gas distributions \cite{Schimd:2024xig}. Significant progress has also been made in deriving analytic expressions for the ensemble averages of MFs for both Gaussian and non-Gaussian random fields \cite{Tomita:1986,Matsubara:2011,Gay:2012,Matsubara:2020,Chingangbam:2023_composite}.

In this work, we quantify the non-Gaussian information contained in LIM maps using MFs, assuming a fiducial astrophysical model. We investigate the prospects for constraining astrophysical parameters through a Fisher-matrix analysis for a given experimental configuration. In this study, we do not account for foregrounds or contamination from interloping spectral lines. We further extend our analysis to a non-standard scenario in which the primordial power spectrum deviates from the standard power-law form through the presence of a localized, bump-like feature. 
We jointly constrain the parameters describing this feature and the astrophysical parameters using the Fisher-matrix formalism. Our results indicate that LIM maps have promising sensitivity to such non-standard primordial signatures. If such a feature exists, LIM observations could potentially provide evidence for its presence; in the absence of a detection, they could instead place stringent constraints on the amplitude and other parameters characterizing the feature.


This paper is structured as follows. In Sec. \ref{sec:LIM_theory}, we briefly explain the theory behind LIM and describe our fiducial astrophysical modelling.  Sec. \ref{sec:inflation} introduces the non-standard, modified inflationary model.
With a brief overview of experimental configuration, we produce LIM maps in Sec. \ref{sec:producing_maps}. We compute MFs for our models and their detectability prospects in Sec. \ref{sec:MF}, \ref{sec:fisher} respectively. We conclude in Sec. \ref{sec:conclusions}. 

\section{Theory and modelling of line intensity mapping}
\label{sec:LIM_theory}
The intensity of observed light within frequency interval $\nu$ and $\nu+{\rm d}\nu$ along a line of sight $\hat n$ is given by,
\begin{equation}
    I_{\nu}(\hat n)=\frac{F_{\nu}(\hat n)}{{\rm d}\Omega{\rm d}\nu},
\end{equation}
where $F_{\nu}$ is the flux, $\nu$ is the observed frequency and ${\rm d}\Omega$ is the solid angle subtended by the observed region in the sky. The observed frequency is related to the rest frame frequency of a particular line emission as,
\begin{equation}
    \nu=\frac{\nu_{\rm rf}}{1+z},
\end{equation}
where $z$ is the redshift of line emitting source. In principle, there can be multiple combinations of $\nu_{\rm rf}$ and $z$ which can contaminate the signal at the observed frequency. In this work, we will not be concerned with this complexity and assume that we can infer $z$ of emitting source, given a combination of $\nu$ and $\nu_{\rm rf}$. The flux is related to the luminosity of line as,
\begin{equation}
    F_{\nu}(\hat n)=\frac{L(\hat n)}{4\pi D_L^2}=\frac{\rho_L(\hat n)}{4\pi D_L^2}\times {\rm d}\Omega D_A^2\times \frac{c(1+z)}{\nu H(z)}{\rm d}\nu, 
\end{equation}
where $\rho_L$ is the comoving luminosity density, $D_L$ is the luminosity distance, $D_A$ is the comoving angular diameter distance with $D_L=(1+z)D_A$ and $H(z)$ is the Hubble rate. In this expression, we have used the fact that the comoving distance along line of sight is ${\rm d}\chi=\frac{c{\rm d}z}{H(z)}$ with $\frac{{\rm d}\chi}{{\rm d}\nu}=\frac{c(1+z)}{\nu H(z)}$. Putting all the factors together, the observed intensity is given by,
\begin{equation}
    I_{\nu}(\hat n)=\frac{F_{\nu}(\hat n)}{{\rm d}\Omega{\rm d\nu}}=\frac{c\rho_L(\hat n)}{4\pi\nu_{\rm rf}H(z)}.
\end{equation}

In order to model $\rho_L$, we use the formalism of halo model \citep{Seljak2000,PS2000}. In principle, one has to do detailed cosmological hydrodynamic simulations including galaxy formation to model the emission which can be computationally prohibitive. Therefore, in literature, semi-numerical computations using halo model, which is computationally much cheaper, is typically used. In this formalism, the dark matter halos provide gravitational potential wells for the baryons to fall into after decoupling from Cosmic microwave Background photons. These baryons form galaxies and give rise to the line emission that we are after. Therefore, the line luminosity as well as the spatial distribution of the emitting sources act somewhat as a proxy of spatial distribution of dark matter. Theoretically, the expression for $\rho_L$ is given as,
\begin{equation}
\rho_L(\hat n, z)=\int {\rm d}M_h L(\hat n, M_h,z)\frac{{\rm d}n(M_h(\vec x),z)}{{\rm d}M_h},
\end{equation}
where $M_h$ is the dark matter halo mass, $L(M_h,z)$ is the  luminosity due to all galaxies inside a dark matter halo with mass $M_h$ and $\frac{{\rm d}n}{{\rm d}M_h}$ is the halo mass function or comoving number density of dark matter halos. Note that the derivative of the halo mass function is wrt the spatially averaged $M_h$. 

In order to produce the spatial distribution of emitting source, we need to resort to simulations. In this work, we have used the semi-numerical simulation {\it 21cmFAST}v4 \citep{MFC2011,MGMMQPW2020} to generate line intensity maps.
We describe the procedure to generate these maps in a forthcoming section. We compute the cumulative luminosity due to all sources within each volume grid of our simulation box. Then the observed intensity is computed as,
\begin{equation}
    I_{\nu}^{i}=\left(\frac{L^i}{\Delta V}\right)\frac{c}{4\pi\nu_{\rm rf}H(z)},
    \label{eq:intensity_sim}
\end{equation}
where $L^i$ is the cumulative luminosity due to all sources within a volume element or voxel indexed by $i$ and $\Delta V$ is volume of each element. 

\subsection{Luminosity model for [C\,{\sc ii}] line}
\label{sec:Luminosity_model}
In this work, we model line intensity maps using [C\,{\sc ii}] line emission as a reference. We note that the same procedure can be applied with other lines. The [C\,{\sc ii}] line (158 $\mu$m, $\nu_{\rm rf}\approx 1900$ GHz) arises from singly ionized carbon atoms with line arising from fine-structure transition between $^2P_{3/2}\rightarrow ^2P_{1/2}$. The [C\,{\sc ii}] line is expected to be a tracer of galaxy or star formation at high redshift. We use a parametric relation obtained in \cite{SSCG2015} to connect luminosity of [C\,{\sc ii}] line to star formation rate (SFR). The expression is given as,
\begin{equation}
    L_{\rm CII}(\hat n, z)=10^{a{\rm log10}({\rm SFR} (\hat n,z))+b}L_{\odot},
    \label{eq:LCII}
\end{equation}
where $a,b$ are assumed to be spatially constant, and  $L_{\odot}$ denotes the luminosity of the sun. The above equation  transforms to 
\begin{equation}
L_{\rm CII}=10^b({\rm SFR})^aL_{\odot}. 
\label{eq:LCII2}
\end{equation}
The SFR is a function of dark matter halo mass $M_h$ and $z$. The authors in \citep{SSCG2015} obtained fiducial values of $a=0.8475$ and $b=7.22$ by fitting to observations at $z\sim 6$, however, there can be some uncertainties in these values. In this work, we use Eq. \ref{eq:LCII} as well as the values of $a$ and $b$ stated above to generate fiducial maps of [C\,{\sc ii}]. Our motivation would be how well the parameters $a$ and $b$ can be constrained using non-Gaussian statistics such as Minkowski functionals, provided we are able to detect this signal with forthcoming experiments. In our result section, we vary $a$ and $b$ and compute the sensitivity of Minkowski functionals to changes in $a$ and $b$. As the final step in modelling, we need an expression for relating star formation rate to dark matter halo mass. We obtain such an expression by fitting a model to galaxy ultraviolet luminosity function (UVLF) data. 

\subsubsection{Modelling SFR with UVLF data}
\label{sec:UVLF_modelling}
The UVLF observable is defined as the comoving number density of galaxies per luminosity at rest frame wavelength of about $1500\AA$. 
The luminosity depends upon the star formation astrophysics inside these galaxies and their redshift evolution is sensitive to cosmological parameters. The UVLF observables have also been used to probe particle nature of dark matter  \citep{DG2024,LKBM2025,DMSL2025,BCLMR2025,BLBMC2025} as well inflationary initial conditions of the universe \citep{YOTT2020,Acharya2026}. The observational data have been made available using Hubble space telescope (HST) and James Webb Telescope (JWST) data at $z\lesssim 15$ \citep{Bouwens2015,Atek2018,Bouwens2021,Donnan2023,Harikane2023,Bouwens2023,McLoed2024}. 


To relate star formation to the mass of dark matter halo, we employ the astrophysical model \cite{SMB2022,Acharya2026} which we describe in brief below. According to this model, the stellar mass within a dark matter halo is related to its mass as,
\begin{equation}
    M_*=f_*(M_h\vec x),z)\left(\frac{\Omega_b}{\Omega_m}\right)M_h(\vec x),
    \label{eq:stellar_mass}
\end{equation}
where $f_*$ is the star formation efficiency which is parameterized as a broken power law,
\begin{equation}
f_*(M_h(\vec x),z)=\frac{\tilde{f}_*}{\left(\frac{M_h(\vec x)}{M_c}\right)^{-\alpha_*}+\left(\frac{M_h(\vec x)}{M_c}\right)^{-\beta_*}}
\end{equation}
The parameters $\tilde{f_*}$ and $M_c$ set the overall efficiency and mass scale of break in power law, respectively. They are assumed to evolve with redshifts as a power law $\tilde{f}_*=\tilde{f}_{*,12}\left[\frac{1+z}{7}\right]^{\epsilon_{\alpha}}$ and $M_c=M_{c,12}\left[\frac{1+z}{7}\right]^{\epsilon_{\beta}}M_{\odot}$ where $\tilde{f}_{*,12}$ and $M_{c,12}$ are respective normalization at $M_h=10^{12}$ $M_{\odot}$. Rest of the parameters do not have any redshift dependence. The star formation rate is expressed as,
\begin{equation}
    {\rm SFR(M_h(\hat n),z)}=\dot{M_*}(M_h(\hat n),z)=\frac{M_*(M_h(\hat n),z)}{t_*(z)},
\end{equation}
where $t_*(z)$ is the timescale of star formation which is assumed to be the Hubble scale or $t_*(z)=1/H(z)$. The galaxy luminosity can be related to its SFR via the scaling relation, 
${\mathcal{K}}_{UV}=\frac{\dot{M_*}(M_h(\hat n,),z)}{L_{UV}}$. The scaling parameter ${\mathcal{K}}_{UV}$ is tuned to observations at low redshifts and with the assumption of Salpeter IMF \citep{MD2014} with a fiducial value of $\mathcal{K}_{UV}=1.15485\times 10^{ -28} M_{\odot}$ ${\rm yr}^{-1}$/ ergs s$^{-1}$ Hz$^{-1}$. Plugging in all the parameters, the UV luminosity is given as,
\begin{equation}
L_{UV}=\frac{f_{*}(M_h,z)}{{\mathcal{K}}_{UV}}H(z)\left(\frac{\Omega_b}{\Omega_m}\right)M_h
\label{eq:L_vs_Mh}
\end{equation}
The galaxy luminosity is converted to magnitude using the relation,
${\rm log_{10}}\left(\frac{L_{UV}}{{\rm ergs}\hspace{0.1cm} {\rm s}^{-1}\hspace{0.1cm}{\rm Hz}^{-1}}\right)=0.4\times (51.6-M_{UV})$.
Finally, the UVLF observable is computed as,
\begin{equation}
\Phi_{UV}(z,M_{UV})=\frac{{\rm d}n}{{\rm d}M_{UV}}=\frac{{\rm d}n(M_h,z)}{{\rm d}M_{h}}\frac{{\rm d}M_h}{{\rm d}L_{UV}}\frac{{\rm d}L_{UV}}{{\rm d}M_{UV}}
\label{eq:Phi_theory}
\end{equation}
where $\frac{{\rm d}n}{{\rm d}M_{h}}$ is the dark matter halo function. The equation above provides an average relation between UVLFs and the magnitude. The authors in \cite{SMB2022} have shown that there is some scatter or stochasticity around this average value which can have important consequences for JWST data at $z\gtrsim 10$
\citep{MCSSMBVQSE2026,SFHSWC2023}. However, for HST data at $z\lesssim 10$, the stochasticity turns out to be un-important  \citep{SMB2022}. We have ignored the effect of stochasticity in this analysis.

The halo mass function is a function of variance of matter density field $\sigma(M_h,z)$ and is given by \citep{PS1974},
\begin{equation}
    \frac{{\rm d}n}{{\rm dln}M_h}=\frac{\rho_m}{M_h}\frac{{\rm dln}\sigma^{-1}}{{\rm dln}M_h}f(\sigma,z).
    \label{eq:HMF_eq}
\end{equation}
In the above equation, $\rho_m$ is the average matter energy density today, while 
 $\sigma^2(M_h)$ is the variance at $z=0$ which is related to the matter power spectrum $P_m$ as,
\begin{equation}
    \sigma^2(R)  =\frac{1}{2\pi^2}\int_0^{\infty}{\rm d}k k^2P_m(k)W^2(kR),
    \label{eq:sigma_R}
\end{equation}
 where $W(kR)$ is top-hat smoothing function in $R$ or position space with $M_h=\frac{4\pi}{3}\rho_mR^3$. We adopt the fitting form of $f(\sigma,z)$ provided by Jenkins et al. \citep{Jenkins2001} with $f(\sigma)=0.315\,{\rm exp}\left(-|{\rm ln}\sigma^{-1}+0.61|^{3.8}\right)$. 
The redshift dependence is captured in $\sigma$ which has the form, $\sigma(M_h,z)=\sigma(M_h)D(z)$, where $D(z)$ is the growth factor which depends upon the assumed cosmological model. We have used the publicly available \texttt{HMF} code \citep{HMF2013} to compute the halo mass function as well as {\it Planck} \cite{Planck2020} best fit cosmological parameters to compute the growth factor. 
We note that some uncertainty may arise from the choice of fitting function. To assess its impact, we repeated the analysis using the fitting function of Tinker et al. \citep{Tinker2008} and found no significant changes in our results. 

Next, we proceed to fit our 6 parameter astrophysical model to the UVLF data. These results are reproduced from \citep{Acharya2026} and briefly explained in Appendix \ref{sec:data}. We plot our posteriors and fit to the data in Fig. \ref{fig:corner_plot} and \ref{fig:fiducial_fit},  respectively. The posterior values of the fit are shown in Table \ref{tab:parameter_table}. In the rest of the work, we use the best fit posterior values in Table \ref{tab:parameter_table} to compute our fiducial star formation model. We use this star formation model in Eq. \ref{eq:LCII} to assign [C\,{\sc ii}] luminosity to dark matter halos with given mass $M_h$.

\begin{table}[!htb]

      \centering
       \begin{tabular}{l|c|r} 
 parameters & prior & posterior values  \\
    \hline
    \vspace{0.1cm}
   $\tilde{f}_{*,12}$ & [0.05,1.0] & $0.571^{0.030}_{-0.034}$ \\ 
   \vspace{0.1cm}
   $M_{c,12}$ & [0.1,10] & $0.351^{0.206}_{-0.138} $ \\
   \vspace{0.1cm}
   $\alpha_*$ & [0,2] & $0.632^{0.151}_{-0.119}$ \\
   \vspace{0.1cm}
   $\beta_*$ & [-2,2] & $-0.557^{0.146}_{-0.170}$ \\
   \vspace{0.1cm}
   $\epsilon_{\alpha}$ & [-5,5] & $0.566^{0.205}_{-0.206}$ \\ \vspace{0.1cm}
   $\epsilon_{\beta}$ & [-5,5] & $-0.449^{0.649}_{-0.616}$ \\
   \hline
 \end{tabular}
 \caption{Prior range for parameters in our star formation model (Sec. \ref{sec:UVLF_modelling}) and their posterior values at 95\% confidence corresponding to the corner plots in Fig. \ref{fig:corner_plot}. These results are reproduced from \citep{Acharya2026}.  }
 \label{tab:parameter_table}
    \end{table}

\section{Modification to the halo mass function due to primordial features}
\label{sec:inflation}
In standard cosmology, the standard inflationary power spectrum is parameterized as a single power law,
\begin{equation}
    P_s(k)=A_s\left(\frac{k}{k_*}\right)^{n_s-1},
\end{equation}
with $k_*=0.05$ Mpc$^{-1}$, $A_s$ and $n_s$ parameterizes the amplitude and slope of power law. This parameterization is motivated by the current Cosmic Microwave Background (CMB) and large scale structure (LSS) measurements within wavenumber $10^{-4}$ Mpc$^{-1}\lesssim k \lesssim 0.1$ Mpc$^{-1}$ \citep{WMAP2011,Planck2020,ACT2025,SPT2025,eBOSS2021,DESI2025}. The parameters are tightly constrained by CMB data \citep{Planck2020} with values ${\rm ln}(10^{10}A_s)=3.044\pm 0.014$, $n_s=0.9649\pm 0.0042$. However, at higher wavenumber, we can have deviation from this standard scenario which have been considered in the past in several works \citep{BH2009,JCGSS2009,JSSS2009,HAJSS2010,MHA2012,CHP2015}.

In this work, we consider a model where a bump-like feature is introduced above the single power spectrum. Bursts of particle production during inflation can generate such features in the power spectrum \citep{CKRT2000,BHKP2009,Furuuchi2016,PPS2017,FNJ2020}. We parameterize the modified power spectrum by including the dominant term as \citep{PPS2017,NCSMF2025},
\begin{equation}
P_s(k)=A_s\left(\frac{k}{k_*}\right)^{n_s-1}+A_I\sum_i\left(\frac{f_1(x_i)}{f_1^{\rm max}}\right),
\label{eq:PPK}
\end{equation}
where $A_I$ is the amplitude of the bump. The function $f_1(x)$ is expressed as,
\begin{equation}
    f_1(x_i)=\frac{[{\rm sin}(x_i)-{\rm SinIntegral}(x_i)]^2}{x_i^3},
\end{equation}
where $x_i=\frac{k}{k_i}$ and $f_1^{\rm max}$ is the maximum value of $f_1(x)$. The parameter $k_i$ is related to the position of $i$th feature which peaks at $k_{{\rm peak},i}=3.35k_i$. We consider one bump or $i=1$ in this work for computational simplicity. 

\begin{figure}
\centering
\includegraphics[width=\columnwidth]{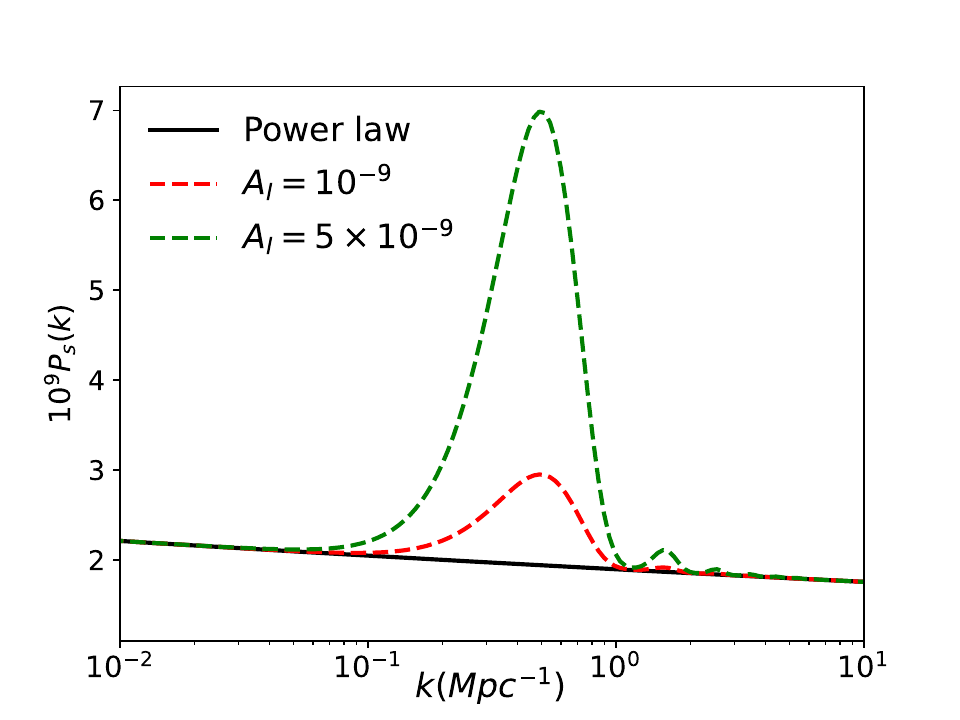}
\caption{Modified power spectrum with a bump-like feature at $k_{\rm peak}=0.5$ Mpc$^{-1}$, $A_I=10^{-9}$ (red) and $A_I=5\times 10^{-9}$ (green). We plot the standard single power law case with best fit parameters $n_s=0.9649$ in black.} 
\label{fig:modified_Ps}
\end{figure}

\begin{figure}[t]
\includegraphics[width=.85\columnwidth]{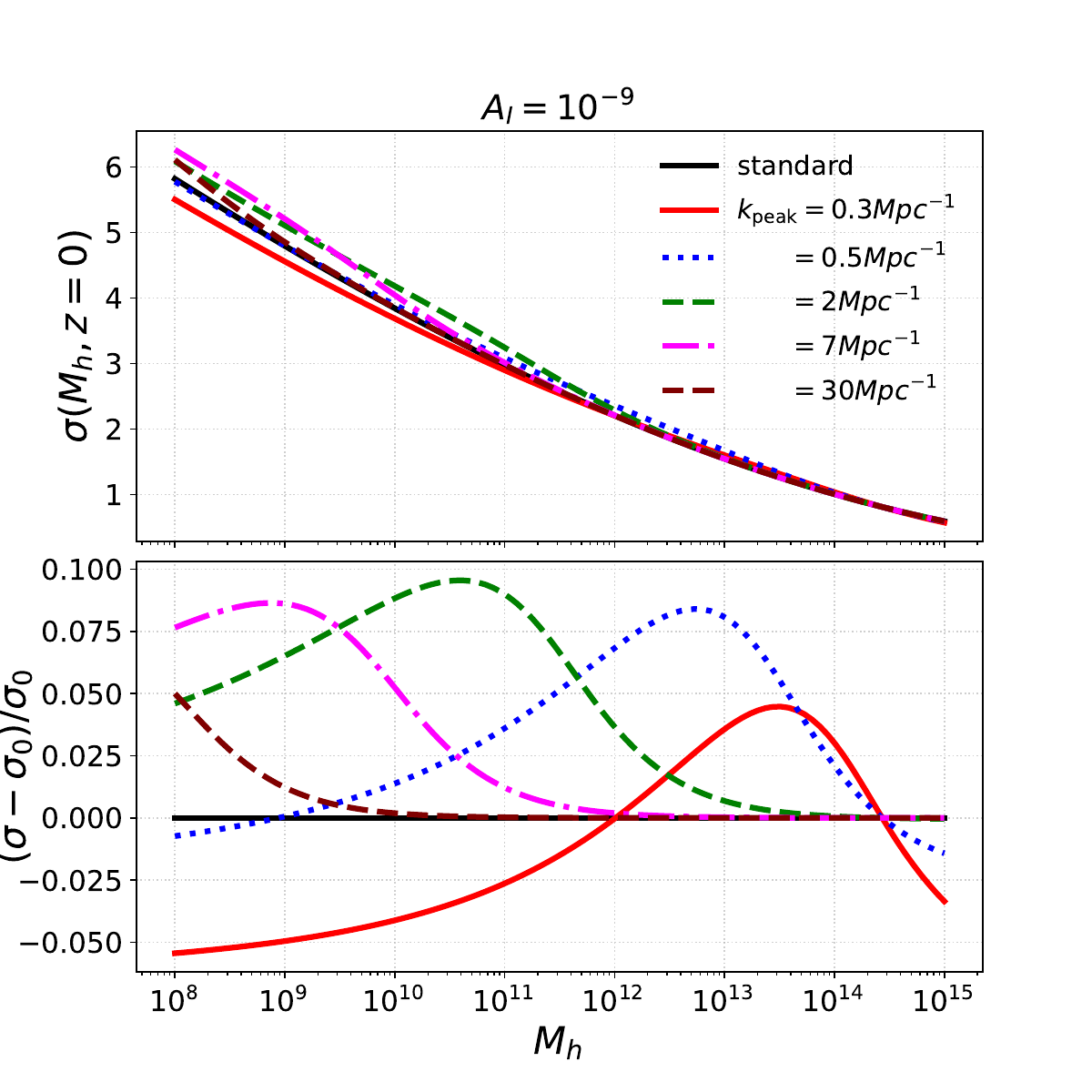}\\
\vskip -.5cm
\includegraphics[width=.85\columnwidth]{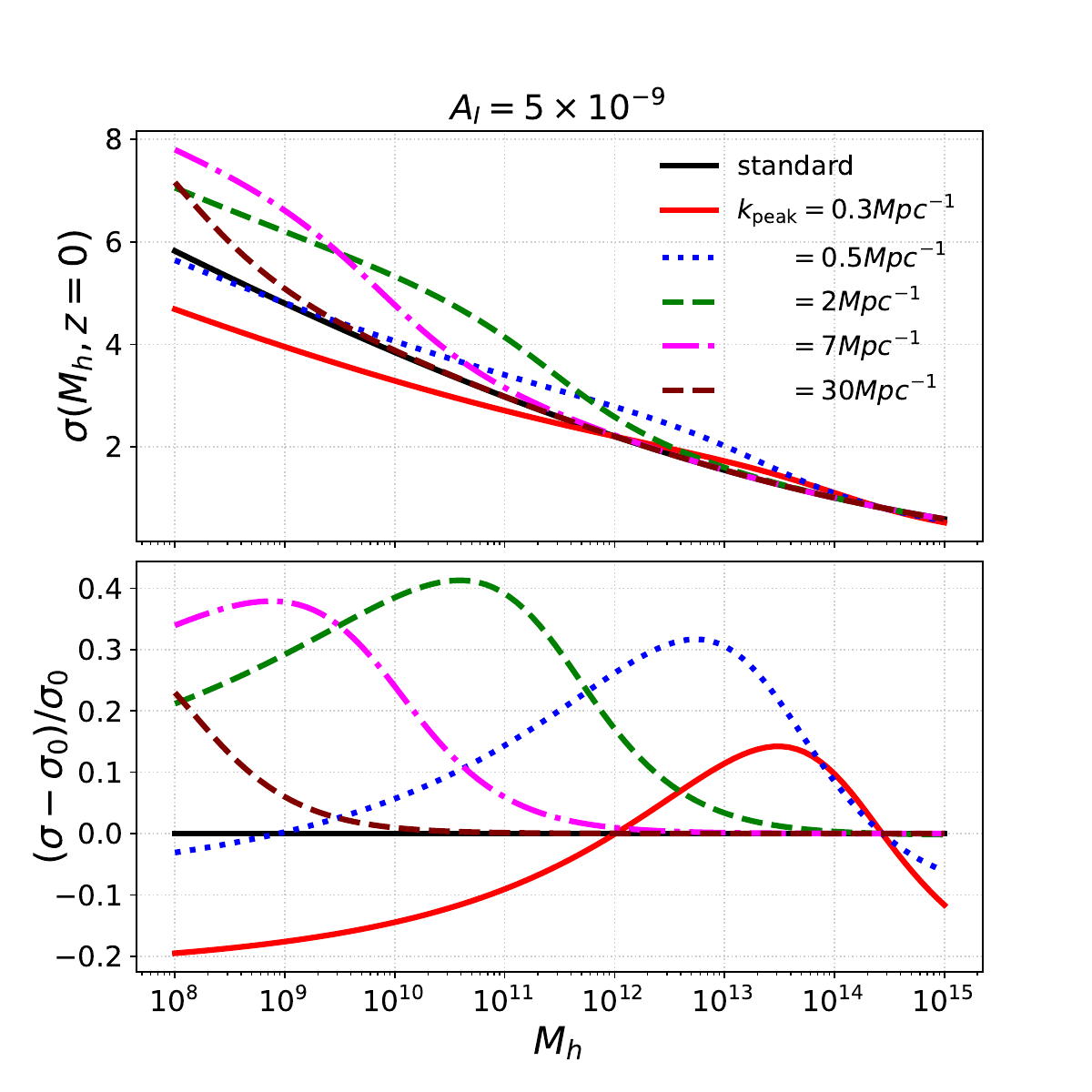}
\caption{$\sigma(M_h)$ and its deviation relative to the standard $A_I=0$ case as a function of halo mass at $z=0$, for different values of $k_{\rm peak}$.  The top panel shows the case of $A_I= 10^{-9}$, while the bottom shows $A_I=5\times 10^{-9}$. We have chosen the normalization $\sigma_8=0.8111$ \citep{Planck2020}.}
\label{fig:modified_sigma}
\end{figure}

In Fig. \ref{fig:modified_Ps}, we plot the modified power spectrum with $k_{\rm peak}=0.5$ Mpc$^{-1}$ and $A_I=10^{-9}$. We plot the corresponding $\sigma(M_h)$ and for a few different combination of $A_I$ and $k_{\rm peak}$ in Fig. \ref{fig:modified_sigma}. The difference between modified $\sigma(M_h)$ and the standard case becomes more significant with increase in $A_I$ as expected. Larger $k_{\rm peak}$ modifies structure formation at progressively smaller length scales or smaller halo mass. We see that the modified $\sigma(M_h)$ falls below the standard case for some range of $M_h$. This is due to the normalization of $\sigma(M_h)$ as denoted in Fig. \ref{fig:modified_sigma} which fixes the total matter content in the universe. Therefore, increase of matter content at some mass scales results in reduction of matter content at different mass ranges. \color{black} Since $f(\sigma)$ (Eq. \ref{eq:HMF_eq}) depends sensitively on $\sigma(M_h)$, any change in $\sigma(M_h)$ shows up in halo mass function (see figure 4  of \cite{NCSMF2025}).  \color{black}

We compute the modified matter power spectrum as $P_m(k)=P_s(k)\mathcal{T}^2(k)$, where $\mathcal{T}(k)$ is the matter transfer function computed using CLASS \citep{BLT2011}. We substitute the matter power spectrum in Eq. \ref{eq:sigma_R} in order to compute $\sigma(M_h)$. We find that $\sigma(M_h)$ changes non-negligibly which will be further magnified in the halo mass function due to its exponential dependence on $\sigma(M_h)$ (Eq. \ref{eq:HMF_eq}). Therefore, the number of dark matter halos is modified and becomes a function of $A_I$ and $k_{\rm peak}$. Since the cumulative [C\,{\sc ii}] luminosity depends upon total number of galaxies which is related to number of dark matter halos, the observed [C\,{\sc ii}] line intensity is also modified. Therefore, we expect to find signature of modification to inflation power spectrum, if present, provided we can distinguish them from our fiducial astrophysical model.

\section{Experimental configuration and producing intensity maps}
\label{sec:producing_maps}
\begin{figure*}[t]
\centering 
{\large{Noiseless \hskip 5.5cm With Gaussian noise }}\\
\includegraphics[width=.9\linewidth]{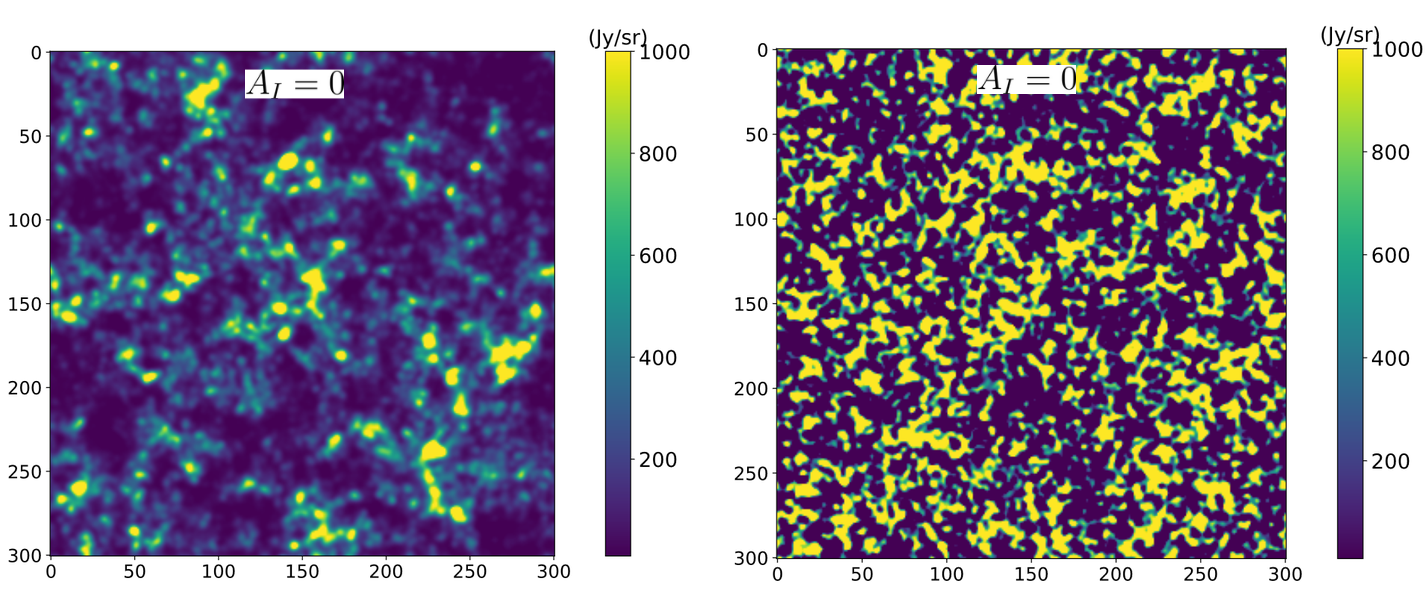}
\includegraphics[width=.9\linewidth]{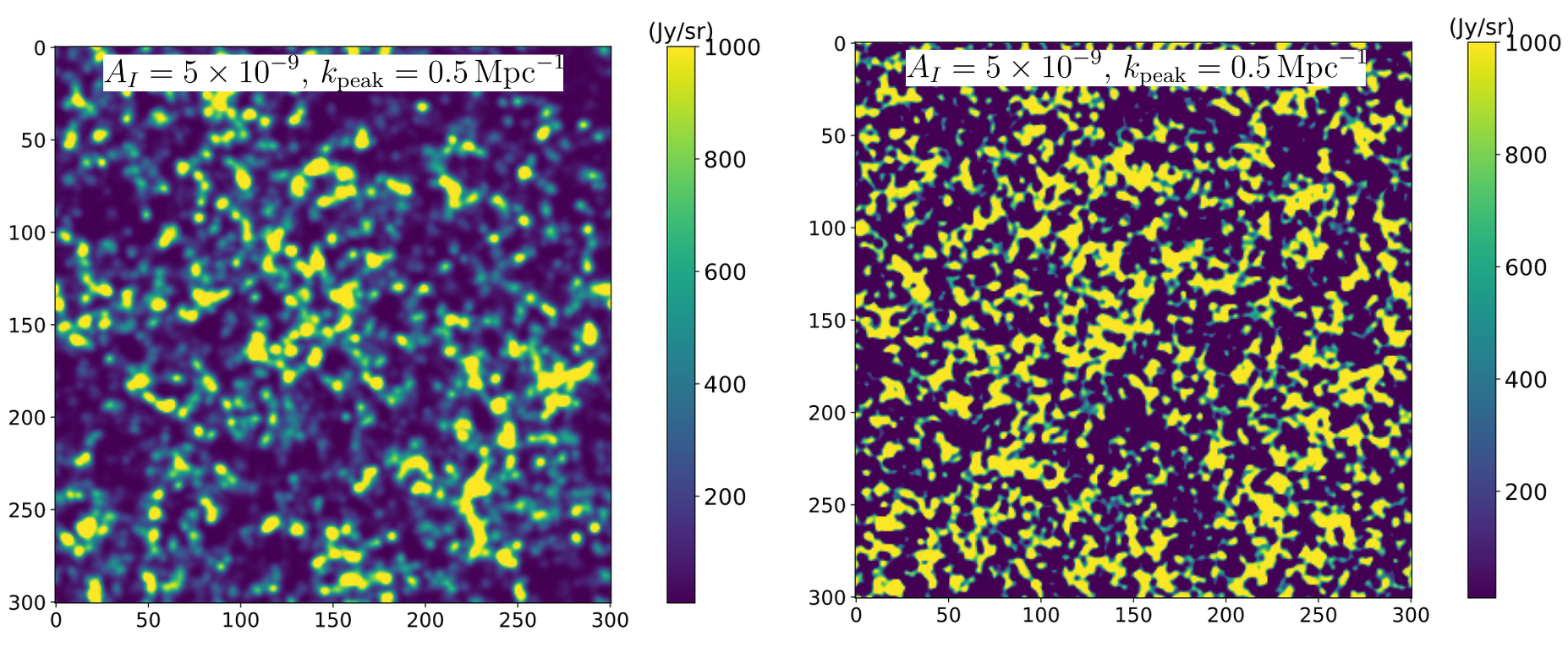}
 \caption{{\it Left column}: [C\,{\sc ii}] intensity maps at $z=5.8$, without noise. The top and bottom panels correspond to $A_I=0$  and $5\times 10^{-9}$, respectively. The smoothing scale used is 1.6 cMpc. The luminosity model is given by Eq. \ref{eq:LCII}, with fiducial parameters $a=0.8475,\ b=7.22$. For the SFR we have used the median values shown in Table. \ref{tab:parameter_table}. {\it Right column}: The same maps as the left ones, but with Gaussian noise added to them and smoothed at  the same scale of  1.6 cMpc. The noise standard deviation used is $\sigma_n=6000$ Jy sr$^{-1}$. This value is considerably higher than the standard deviations of the ideal signals which are 255 and 380 Jy sr$^{-1}$, respectively, and hence the images appear to be overwhelmed by noise. }
    \label{fig:Intensity_grid}
\end{figure*}

We consider the experimental configuration of the Cerro Chajnantor Atacama Telescope (\texttt{CCAT}) collaboration \citep{CCAT} for generating [C\,{\sc ii}] intensity maps. The \texttt{CCAT} deep spectroscopic survey (DSS) covers a field of view of 4 deg$^2$ and has an observing time of 2000 hours, with four central frequency channels centred at 220, 280, 350 and 410 GHz  (Table 1 of \citealt{CCAT}).

In this work, we focus on two frequency bands. The first is centered at $\sim 280$ GHz, which corresponds to [C\,{\sc ii}] emission from sources at $z \sim 5.8$. The frequency resolution is approximately $\Delta \nu = 2.8$ GHz, corresponding to a redshift resolution of $ \Delta z = \frac{\Delta \nu}{\nu_{\rm rf}}(1+z)^2 \approx 0.07$. 
at $z = 5.8$, where $\nu_{\rm rf}$ is the rest-frame frequency of the [C\,{\sc ii}] line. This translates into a line-of-sight spatial resolution of $ \approx c\Delta z/H(z) \approx 29 \, {\rm cMpc}$.
The beam full width at half maximum (FWHM) in this frequency band is 48 arcsec. Assuming a Gaussian beam, the corresponding angular resolution is $\theta_{\rm res} = \mathrm{FWHM}/\sqrt{8\ln 2} \approx 20 \, {\rm arcsec}$. 
At $z = 5.8$, this corresponds to a transverse spatial resolution of approximately 0.82 cMpc. Finally, the 2$^\circ$ angular extent of the survey corresponds to a transverse comoving scale of approximately 283 cMpc. Accordingly, for our simulations, we adopt a box size of 300 cMpc, discretized into $300^3$ grid cells.

The second frequency band we focus on is centered at $\sim 220$ GHz, which corresponds to [C\,{\sc ii}] emission from sources at $z \sim 7.6$. 
The frequency resolution is of $\Delta \nu = 2.2$ GHz then corresponds to a redshift resolution of $ \Delta z=0.09$ and spatial resolution along the line of sight of $\approx$ 25 cMpc. The FWHM is 58 arcsec, which corresponds to resolution in transverse directions of $\approx$ 1.07 cMpc. 
Again, for simulations at this redshift, we adopt a box size of 300 cMpc, discretized into $300^3$ grid cells.

{\bf Noise maps at $z=5.8$ and $7.6$}: We simulate the noise maps at these two redshifts as zero mean 2D Gaussian fields.  Following Table 1 of \cite{CVCW2020}, the noise standard deviation for $z=5.8$ is taken to be $\sigma_n\approx$ 6000 Jy sr$^{-1}$, while for  $z=7.6$ it is taken to be $\sigma_n\approx$ 4000 Jy sr$^{-1}$.

\subsection{Simulating ideal [C\,{\sc ii}] LIM maps and the impact of noise addition}


We use the halo sampler of \texttt{21cmFASTv4} \citep{DMM2025} to generate dark matter halo catalogs.  \texttt{21cmFAST} does not explicitly solve for the gravitational evolution of dark matter particles. Instead, it employs the excursion-set formalism to identify regions of enhanced density that can be associated with dark matter halos. The gravitational evolution of these halos is approximated using Lagrangian perturbation theory, which displaces them from their initial positions to capture the effects of gravitational evolution and reproduce the large-scale clustering of halos. This approach makes \texttt{21cmFAST} computationally much faster than traditional $N$-body simulations. 
However, these approximations do not fully resolve the internal matter distribution or the detailed spatial structure of individual halos. 
At a spatial resolution of $\sim 1$ cMpc, the internal distributions of dark matter and baryons within individual halos are generally unresolved. We therefore treat each dark matter halo as a spatially unresolved source of line emission, effectively representing it as a point-like emitter with no resolved internal spatial structure.

We consider dark matter halos with masses $M_h \gtrsim 10^{10}\,M_{\odot}$ as potential hosts of [C\,{\sc ii}]-emitting sources. Navarro et al. \citep{Navarro2018} investigated the relationship between neutral hydrogen and its host dark matter halos using cosmological hydrodynamical simulations and found an exponential cutoff in the halo population toward lower masses, around $M_h \sim 10^{10}\,M_{\odot}$. We therefore adopt $10^{10}\,M_{\odot}$ as the minimum halo mass, consistent with previous studies \citep{HCBL2024}. For each identified halo, we use its mass $M_h$ and Eq.~\ref{eq:LCII} to assign the corresponding [C\,{\sc ii}] luminosity. We then sum the luminosities of all sources within each grid cell to obtain the total [C\,{\sc ii}] luminosity per cell.

For computational simplicity, we restrict our analysis to two-dimensional (2D) intensity maps rather than constructing a full three-dimensional (3D) intensity map. Given the relatively small angular field of view of $\sim 2^\circ$, we adopt the plane-parallel approximation and neglect the curvature of the sky. Owing to the finite frequency resolution of the instrument, individual 2D slices of thickness 1 cMpc (the grid-cell size in our simulations) cannot be resolved along the line of sight. To account for this limitation, we sum the contributions from all sources across 29 consecutive slices, corresponding to a line-of-sight depth of 29 cMpc, to construct a 2D intensity map at $z=5.8$. 
For carrying out the morphological analysis, we smooth the map in the transverse direction using a Gaussian kernel with at three resolutions 1.6, 3 and 6 cMpc, roughly corresponding to angular scales of $40$,72,144 arcsec, respectively.



For simulating  [C\,{\sc ii}] LIM maps at $z=7.6$ we follow similar steps with some numerical adjustments. The spatial resolution in this case corresponds to $\approx 25$ cMpc and, therefore, $\Delta V=25$ cMpc$^3$. For calculations involving this particular redshift, we have used Gaussian smoothing scale of 3 cMpc which turns out to be resolution of $\approx 70$ arcsec.

To incorporate the primordial power spectrum feature introduced in Sec. \ref{sec:inflation},  we modified the input primordial power spectrum in \texttt{21cmFAST} in the form given by Eq. \ref{eq:PPK}. By varying the two parameters $A_I$ and $k_{\rm peak}$ one can produce  [C\,{\sc ii}]  maps of varying amplitude and location of the primordial feature.

The left panels of figure \ref{fig:Intensity_grid} illustrate the effect of the modified halo mass function on the [C\,{\sc ii}] intensity maps. The top left panel  shows the simulated 2D intensity map for the fiducial case without  instrumental noise. The bottom left panel shows the corresponding map for the same random seed, but for $A_I = 5\times10^{-9}$ and $k_{\rm peak}=0.5,{\rm Mpc}^{-1}$. As discussed previously, a non-zero value of $A_I$ enhances $\sigma(M_h)$ by increasing the abundance of dark matter halos over specific halo-mass ranges determined by $k_{\rm peak}$. For $k_{\rm peak}=0.5,{\rm Mpc}^{-1}$, halos in the rough range 
$M_h\in[10^{10},10^{14}]\,M_{\odot}$ are most strongly affected (Fig. \ref{fig:modified_sigma}). The enhanced abundance of these halos consequently increases the SFR within them and, hence, their [C\,{\sc ii}] luminosity. 
At redshifts around $z\sim 6$ being considered in this paper, such massive halos are relatively rare and highly clustered, resulting in a distinctly non-Gaussian spatial distribution of the [C\,{\sc ii}] intensity. This is manifested as localized patches of high-intensity emission in the maps. As $A_I$ is increased, the abundance and prominence of these luminous regions increase, as can be seen in the left panels of Fig. \ref{fig:Intensity_grid}. 

The right panels of Fig. \ref{fig:Intensity_grid} illustrate the impact of Gaussian instrumental noise on the corresponding ideal signal maps shown in the left panels. For a consistent comparison, the signal-plus-noise maps are smoothed at the same scale of 1.6 cMpc.
 The value $\sigma_n=6000$ Jy sr$^{-1}$ is considerably higher than the standard deviations of the ideal signals which are 255 (top left) and 380 (bottom left)  Jy sr$^{-1}$. Consequently, the resulting maps appear to be dominated by noise. Nevertheless, the underlying signals remain visually discernible in regions of high intensity. The impact of instrumental noise in the subsequent analysis will be further suppressed at larger smoothing scales. 
 

\begin{figure}[b] 
\centering
Without noise \hskip 2.3cm With noise\\
\includegraphics[width=1\linewidth,height=2.1in]{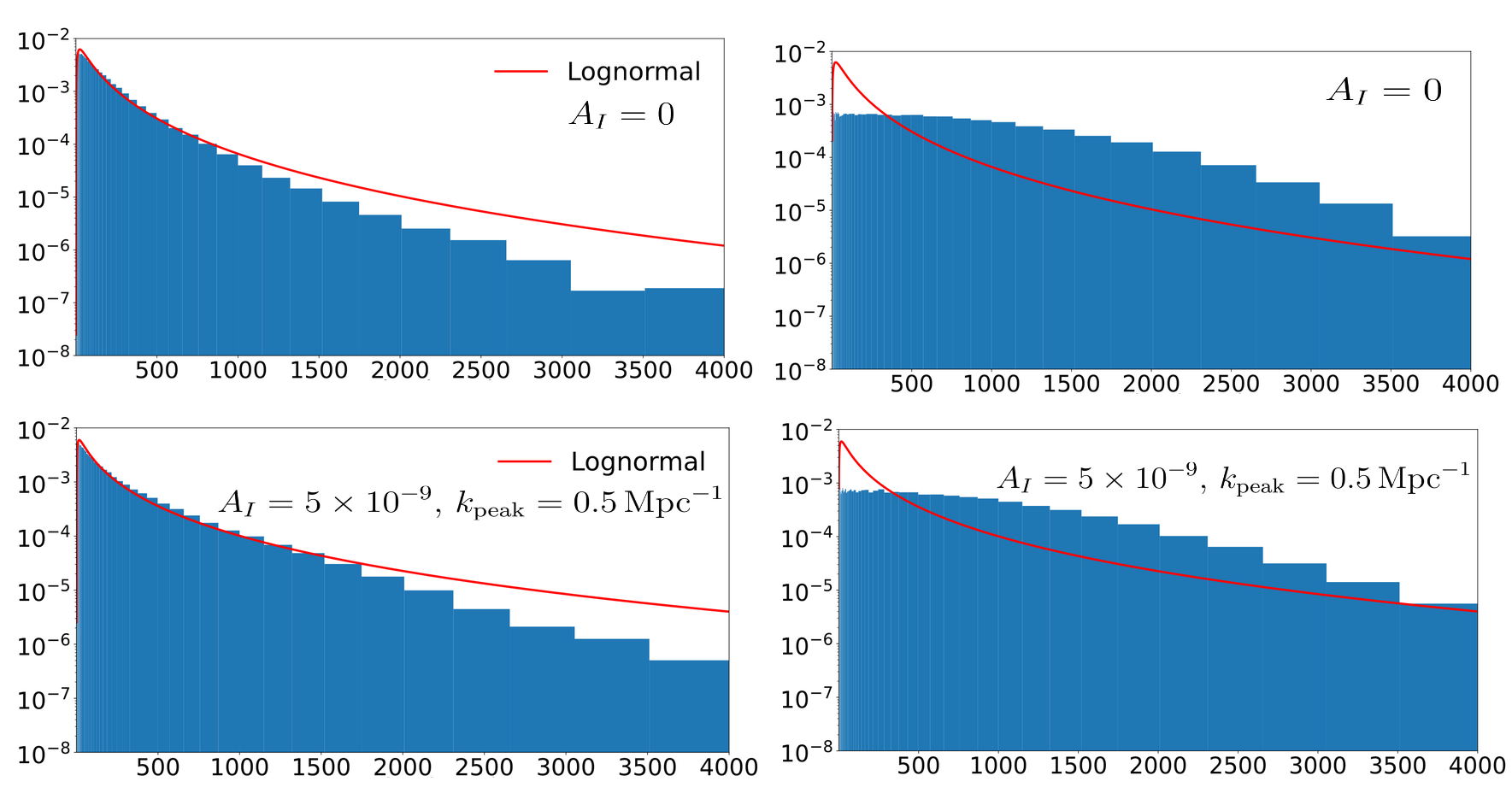}\\ 
{\small{Intensity (Jy/sr) \hskip 2cm Intensity (Jy/sr) }}
                   \caption{Histograms of [C\,{\sc ii}] intensity maps for the same cases as in Fig. \ref{fig:Intensity_grid}. The red lines show the best-fit lognormal PDF curves. }
                   \label{fig:histogram}
\end{figure}


Before examining the morphological properties of the [C\,{\sc ii}] line-intensity maps, we first characterize their one-point probability distribution functions (PDFs). The PDFs provide a simple diagnostic of the non-Gaussianity discussed above and help anticipate how the underlying changes in the intensity distribution may manifest in the morphology of the maps. Figure \ref{fig:histogram} shows the corresponding intensity histograms for the maps presented in Fig. \ref{fig:Intensity_grid}, in the same order.
 The red curves denote the best-fit lognormal forms. For the noiseless featureless fiducial model, the histogram exhibits a departure from lognormality at higher field values. In the bottom left panel, we see that increasing $A_I$ causes the field to approach a lognormal distribution over progressively higher field values. This behavior quantifies 
 the enhancement of $\sigma(M_h)$ induced by the bump feature, which leads to an increased SFR, and consequently, enhanced [C\,{\sc ii}] luminosity over a range of halo masses determined by $k_{\rm peak}$, seen in Fig. \ref{fig:Intensity_grid}. 
  The right panels demonstrate that the addition of Gaussian noise drives the field towards  Gaussian forms, as mentioned earlier.

\section{Minkowski functionals of [C\,{\sc ii}] line intensity maps}
\label{sec:MF}

In this section, we quantify the non-Gaussian information in LIM maps using Minkowski functionals (MFs). We first briefly introduce  Minkowski functionals, and then discuss them for [C\,{\sc ii}] LIM maps. 

\subsection{Overview of Minkowski functionals}
\label{sec:MF_A}

Let $f(x)$ denote a mean subtracted random field, having standard deviation $\sigma_0$, defined on a manifold $M$, with $x$ being coordinates on $M$. The set of all points such that $f(x)\ge\nu\sigma_0$ forms the excursion set $Q_{\nu}$ with boundary $\partial Q_{\nu}$. We refer to $\nu$ as {\it threshold}. 
In 2D, the geometry and topology of $Q_{\nu}$ can be captured by 3 MFs which are defined as \citep{Tomita1986,GPJBBBS1990,MBW1994,SG1998},
\begin{eqnarray}
    V_0(\nu)&=&\frac{1}{A}\int_{Q_{\nu}}{\rm d}a,\\
     V_1(\nu)&=&\frac{1}{4A}\int_{\partial Q_{\nu}}{\rm d}s,\\
     V_2(\nu)&=&\frac{1}{2\pi A}\int_{\partial Q_{\nu}}\kappa {\rm d}s, 
    \label{eq:MF_eqn}
\end{eqnarray}
where $A$ is the observed sky survey area. The first MF, $V_0(\nu)$, is the normalized total area of $Q_\nu$. The other two MFs capture the perimeter and the difference between the counts of connected regions and holes, 
respectively. For zero mean Gaussian fields, the ensemble expectations of the MFs have the following simple forms \citep{Tomita1986},
\begin{eqnarray}
    V_0(\nu)&=&\frac{1}{2}{\rm erfc}\left(\frac{\nu}{\sqrt{2}}\right),\\
    V_1(\nu)&=&\frac{1}{8\sqrt{2}}\left(\frac{\sigma_1}{\sigma_0}\right){\rm e}^{-\nu^2/2},\\
    V_2(\nu)&=&\frac{1}{2(2\pi)^{3/2}}\left(\frac{\sigma_1}{\sigma_0}\right)^2\nu{\rm e}^{-\nu^2/2},
    \label{eq:MF_gaussian}
\end{eqnarray}
where $\sigma_1=\sqrt{\langle|\nabla f|)^2\rangle }$.
\begin{figure*}[t!] 
\includegraphics[height=1.7in]{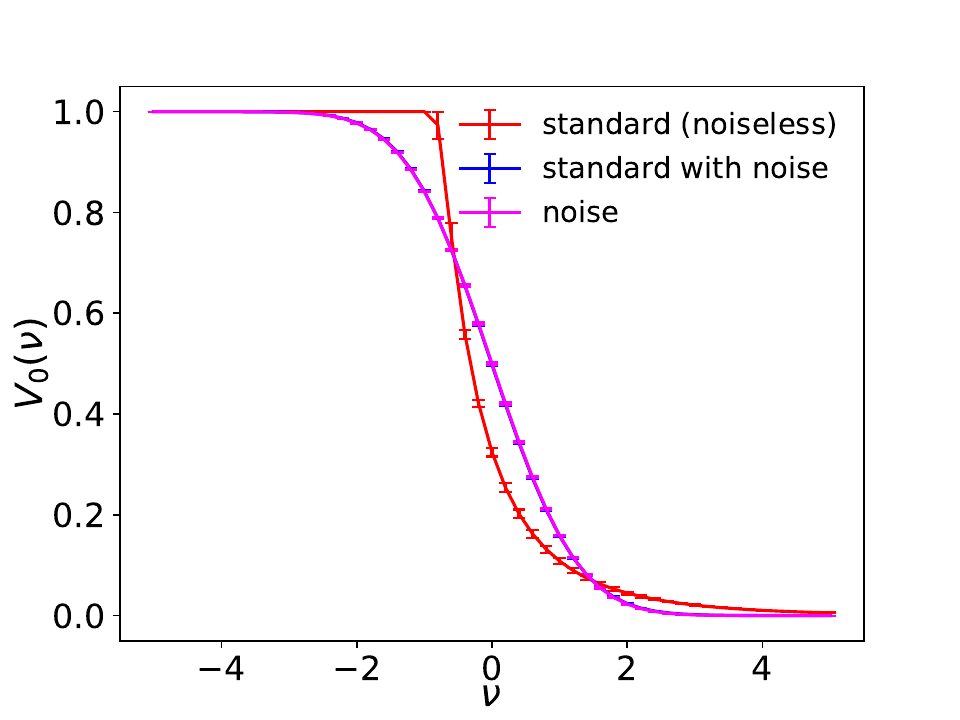}
\includegraphics[height=1.7in]{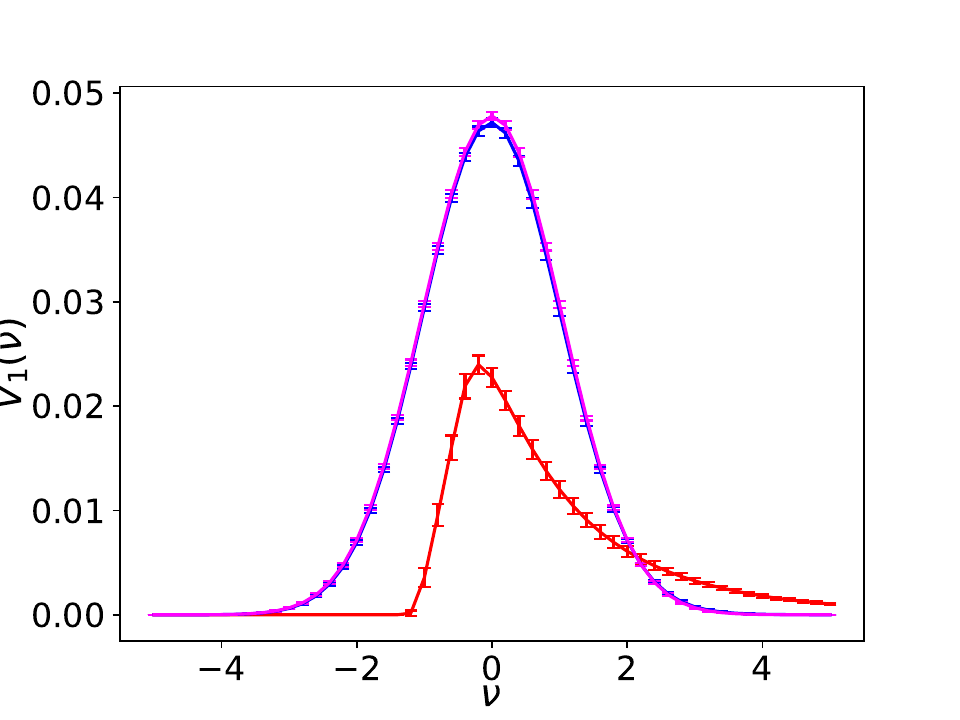}
\includegraphics[height=1.7in]{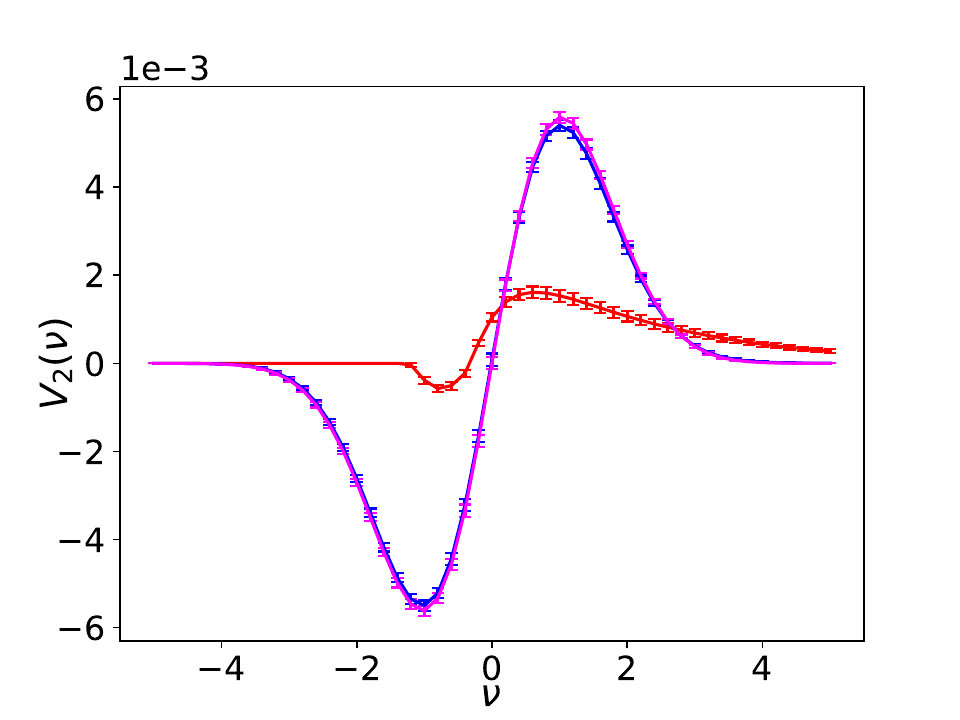}
    \caption{MFs for the LIM map for the fiducial model ($a=0.8475,\, b=7.22,\, A_I=0$), both without and with noise ($\sigma_n=6000$ Jy sr$^{-1}$), and for the pure noise map. The error bars are the standard deviations obtained from 150 realizations. The smoothing scale used is 1.6 cMpc.}
    \label{fig:MF_std_with_noise}
\end{figure*}

\begin{figure*}[t!] 
\centering
{\bf (A) Effects of varying $a,b$:  ideal signal (top row) and signal-plus-noise (lower row)}\\
 \includegraphics[height=2.0in]{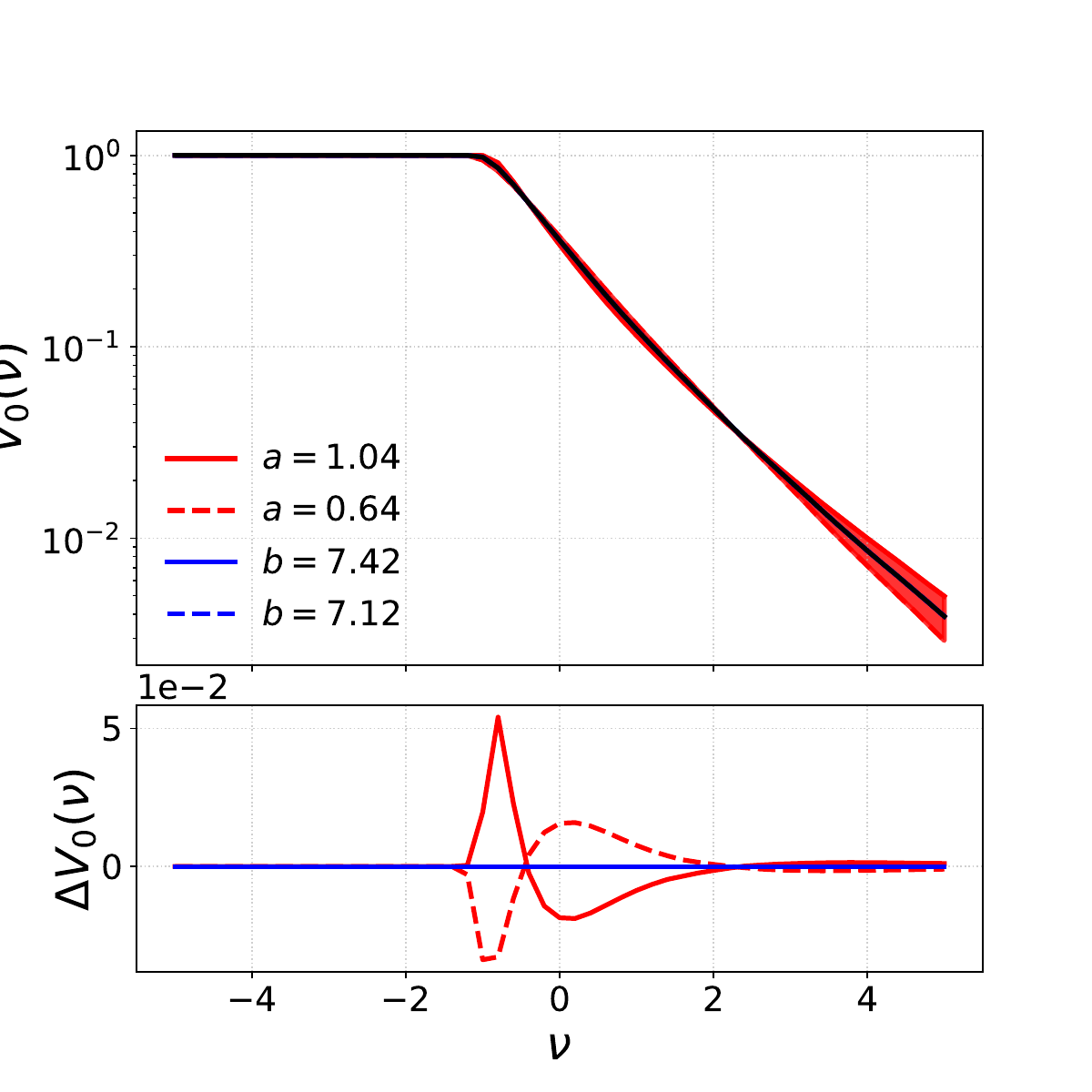}
 \includegraphics[height=2.0in]{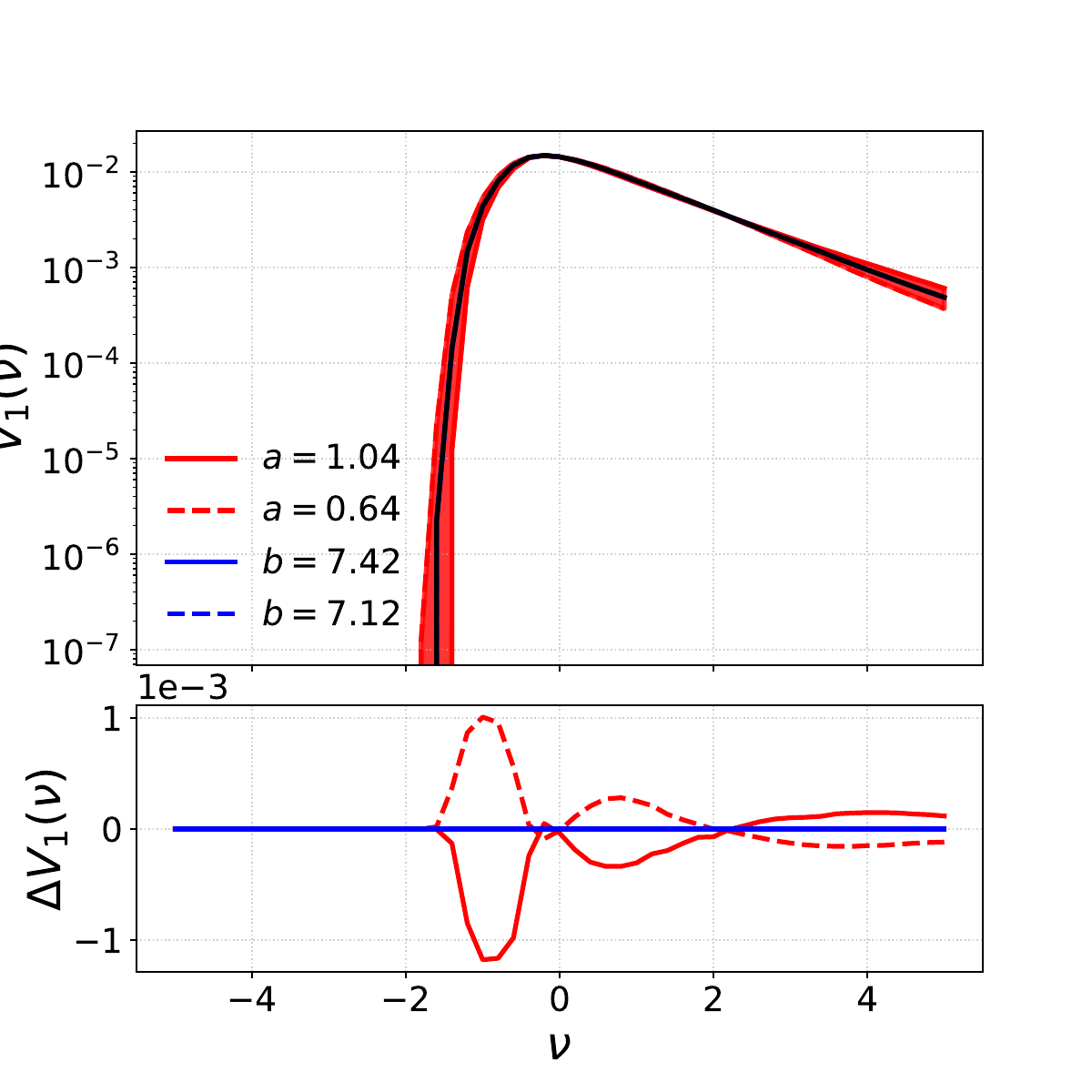}
 \includegraphics[height=2.0in]{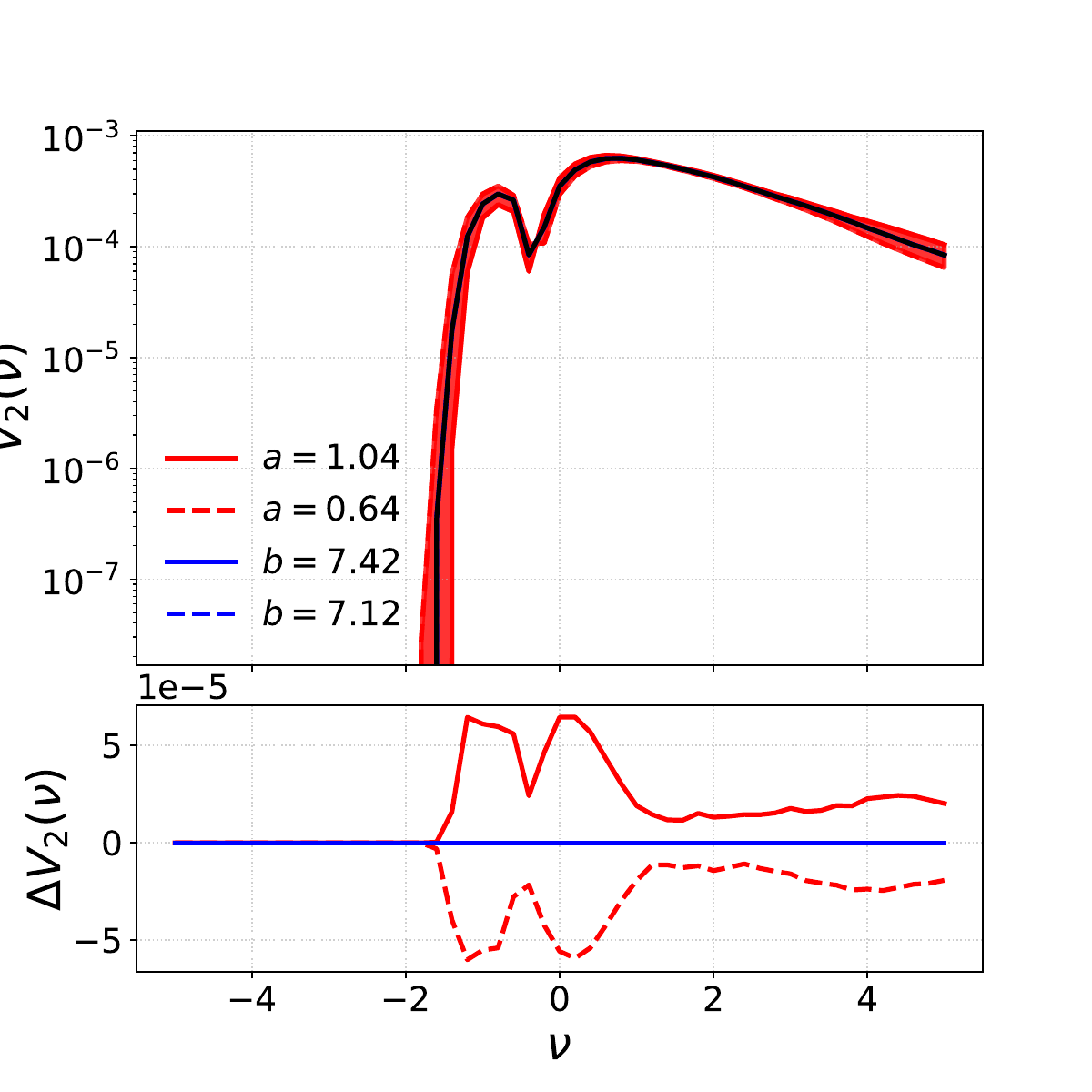}\\
\vskip -0.2cm
 \includegraphics[height=2.0in]{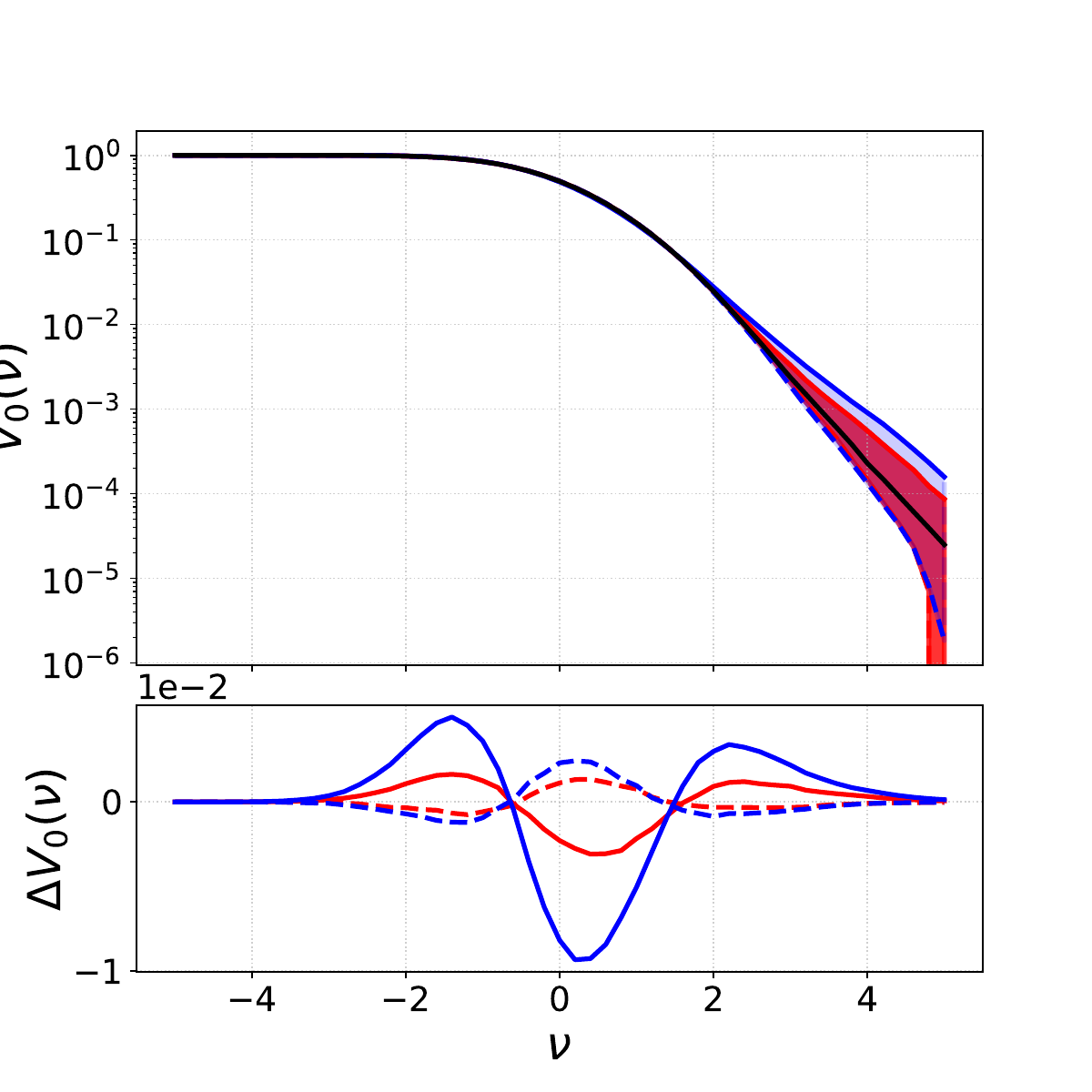}
 \includegraphics[height=2.0in]{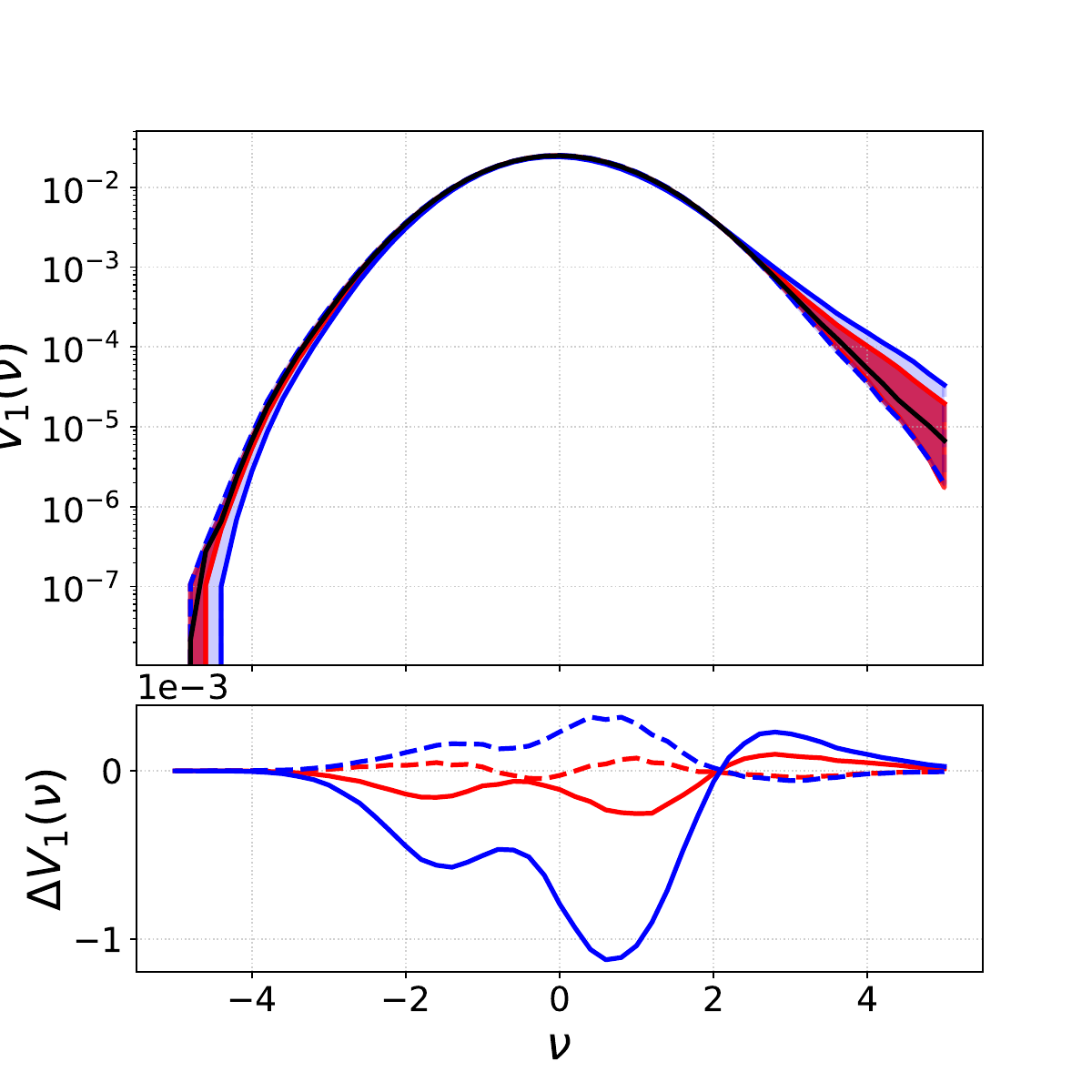}
 \includegraphics[height=2.0in]{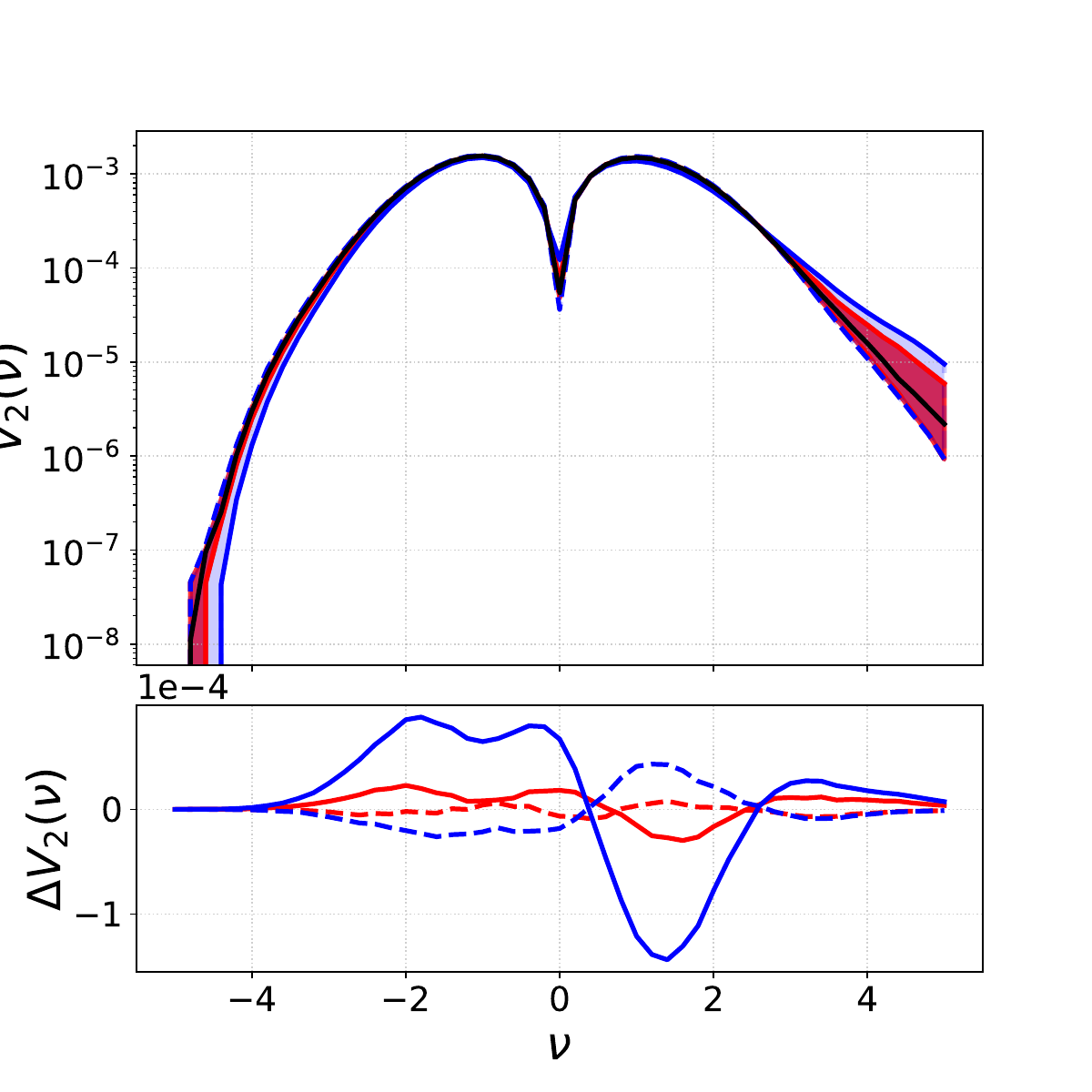}\\
 {\bf (B) Effects of varying of $A_I,\, k_{\rm peak}$: ideal signal (top row) and signal-plus-noise (lower row)}\\
 \includegraphics[height=2.0in]{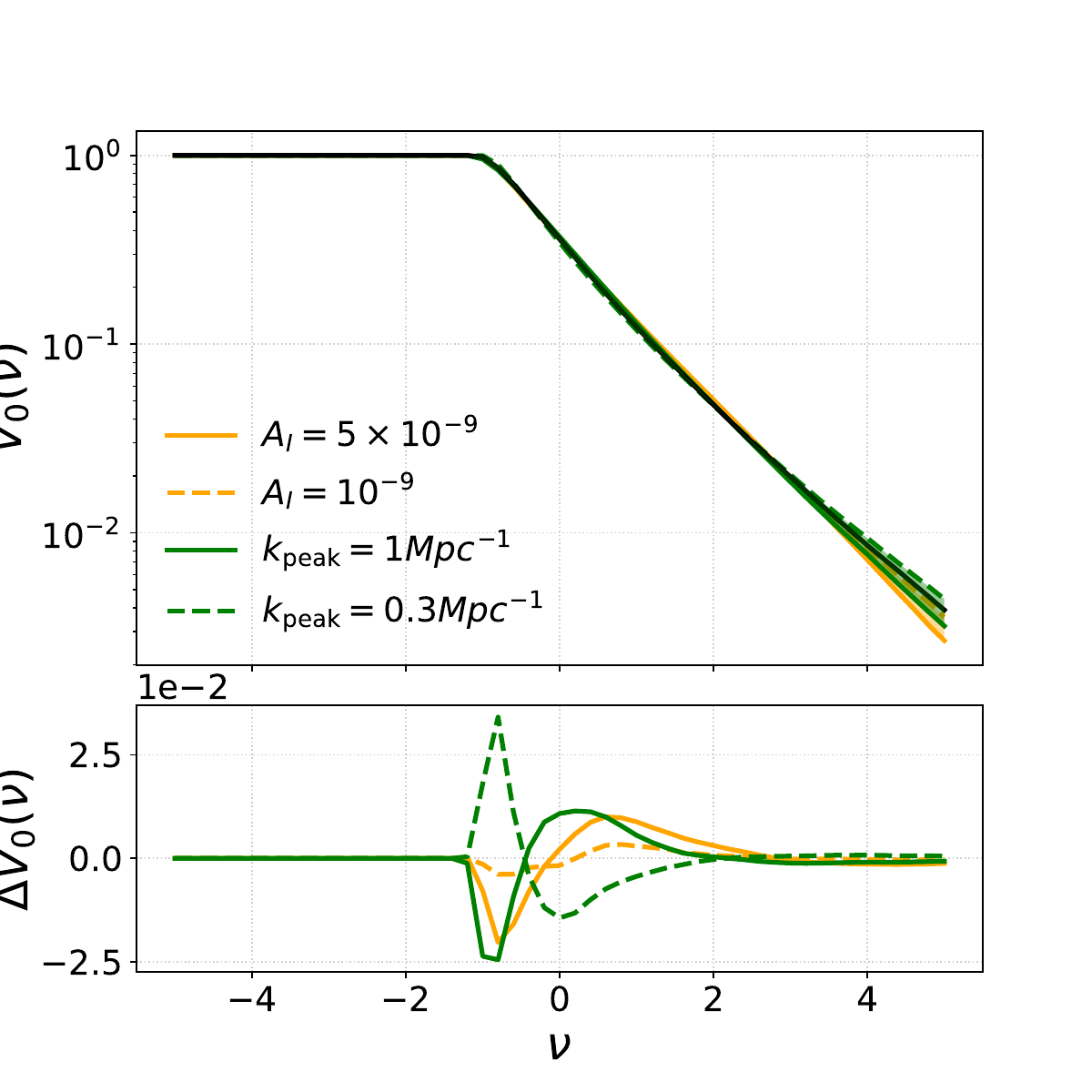}
 \includegraphics[height=2.0in]{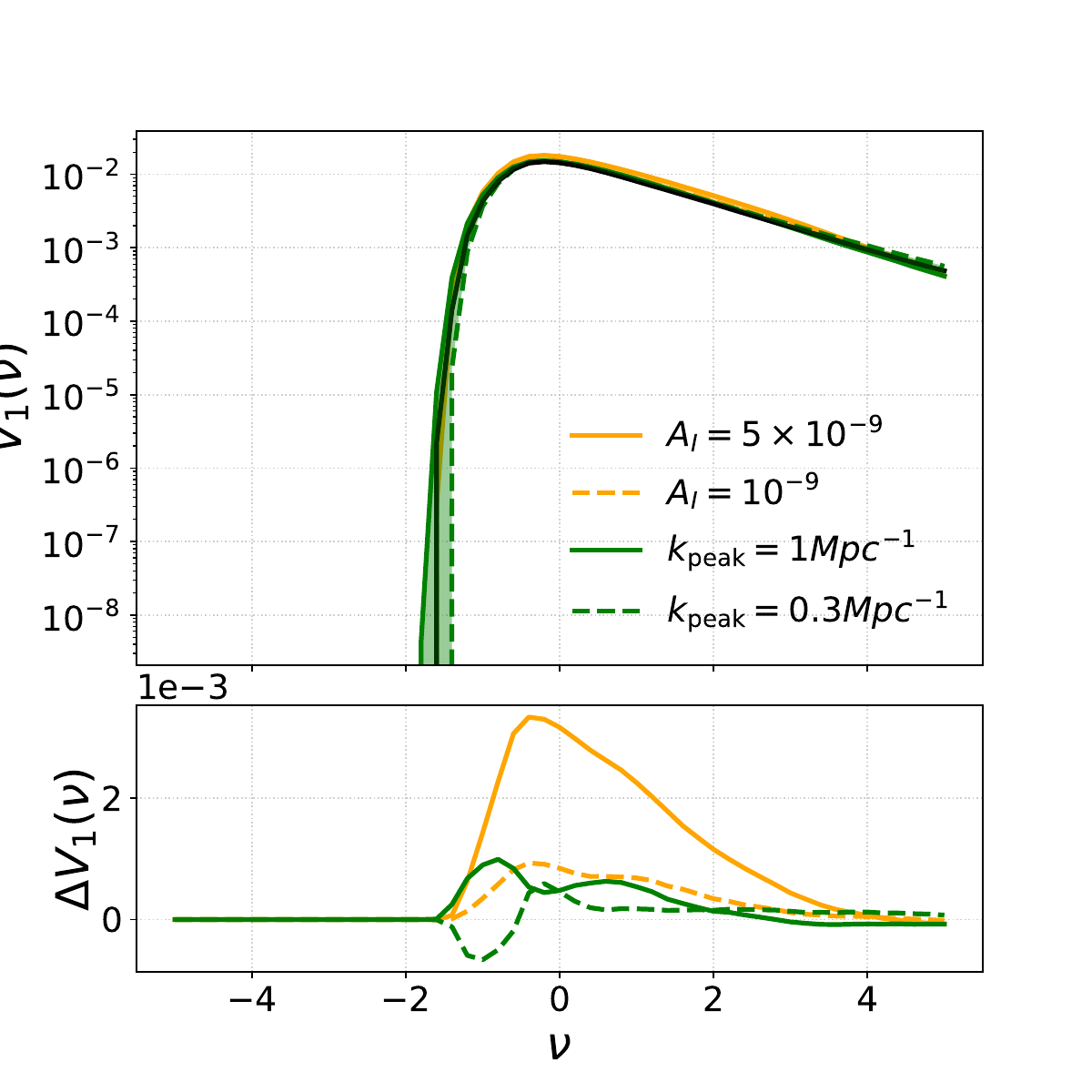}
 \includegraphics[height=2.0in]{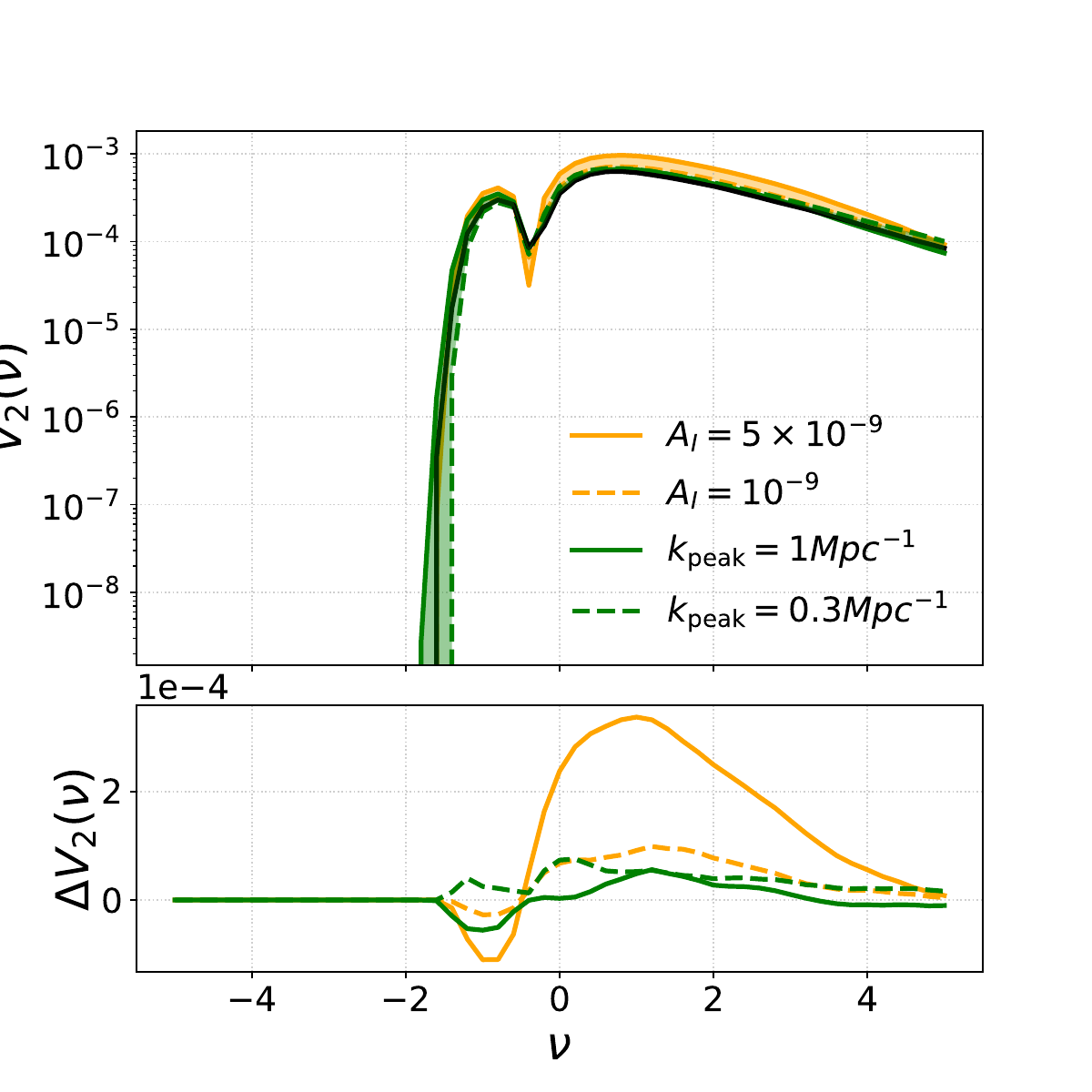}\\
 \includegraphics[height=2.0in]{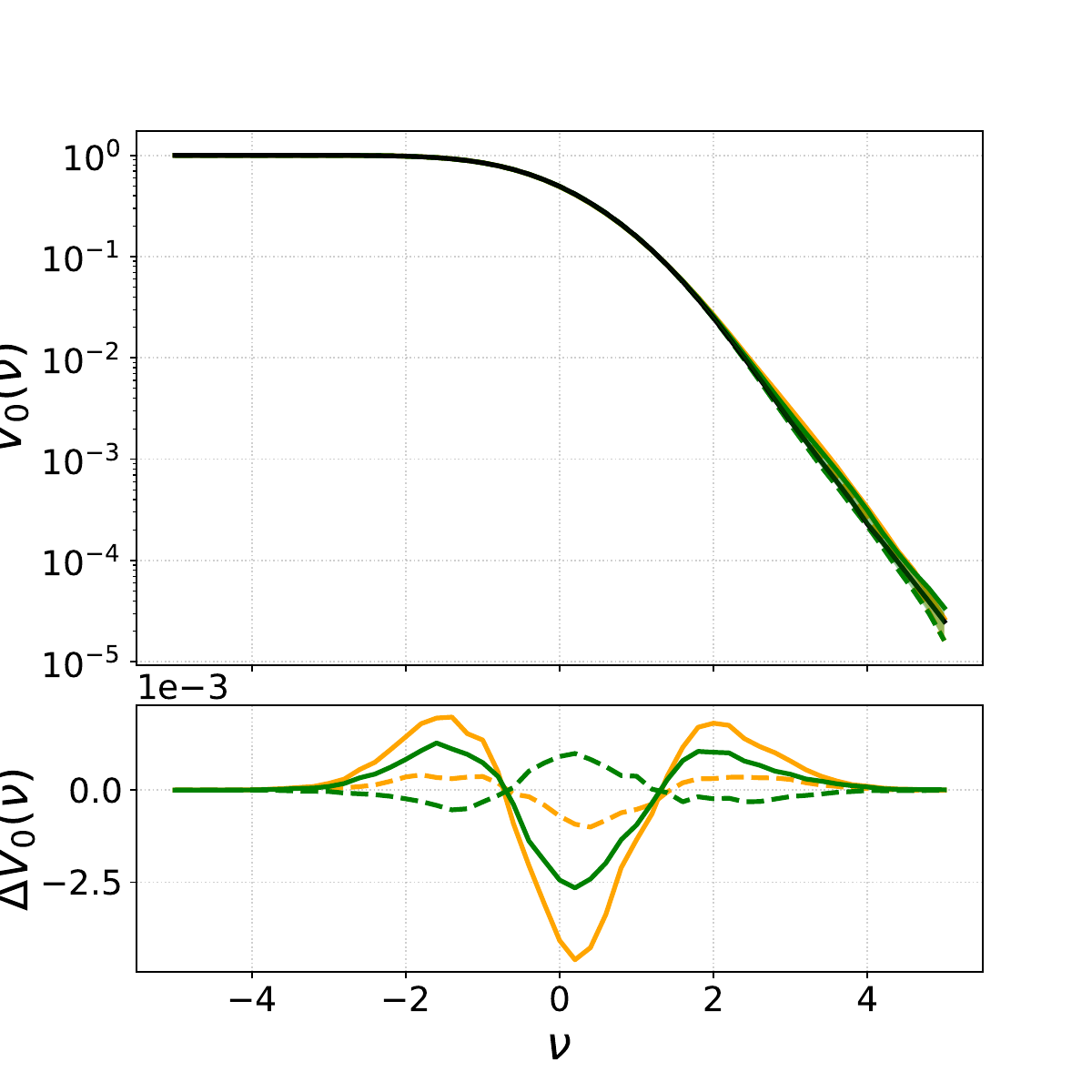}
 \includegraphics[height=2.0in]{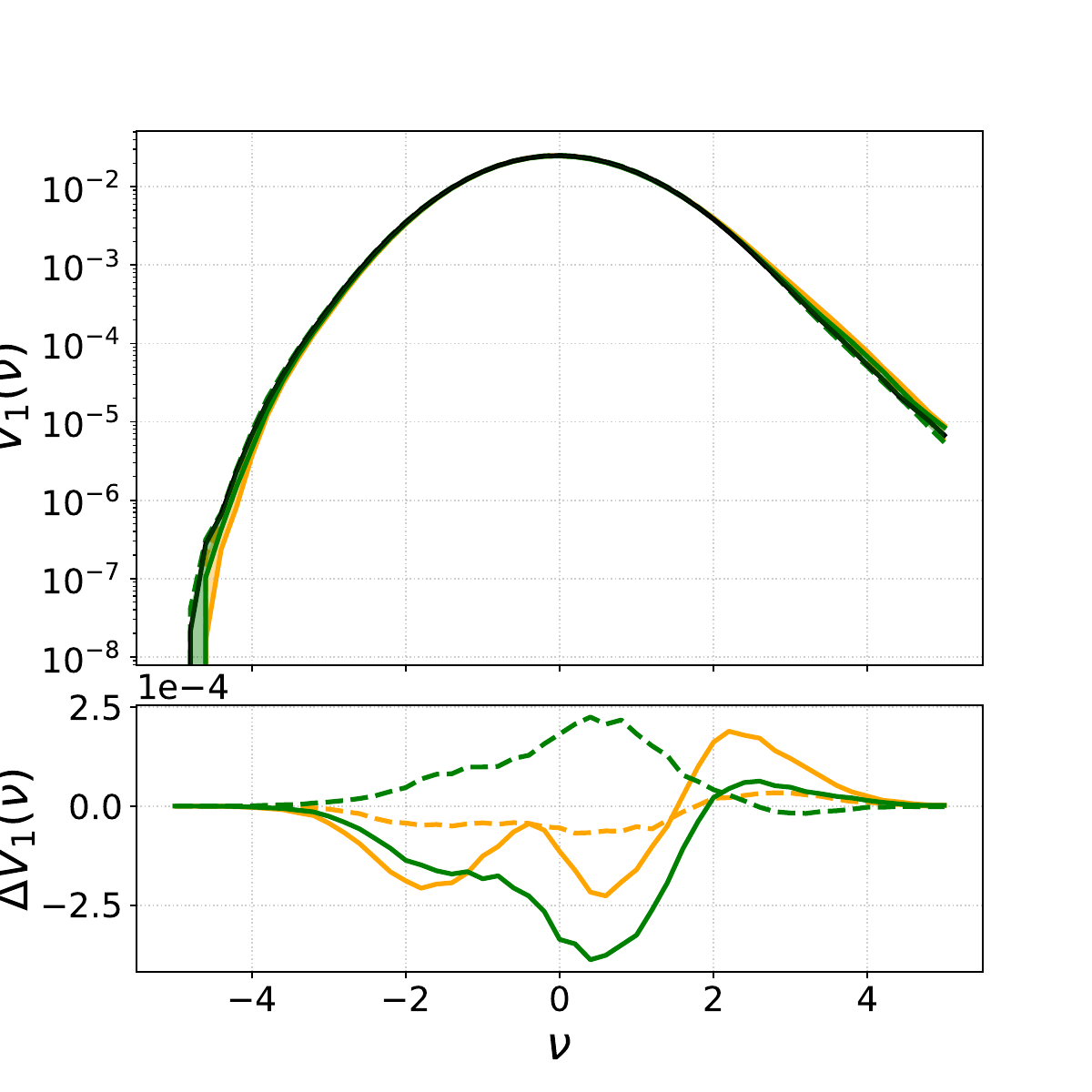}
 \includegraphics[height=2.0in]{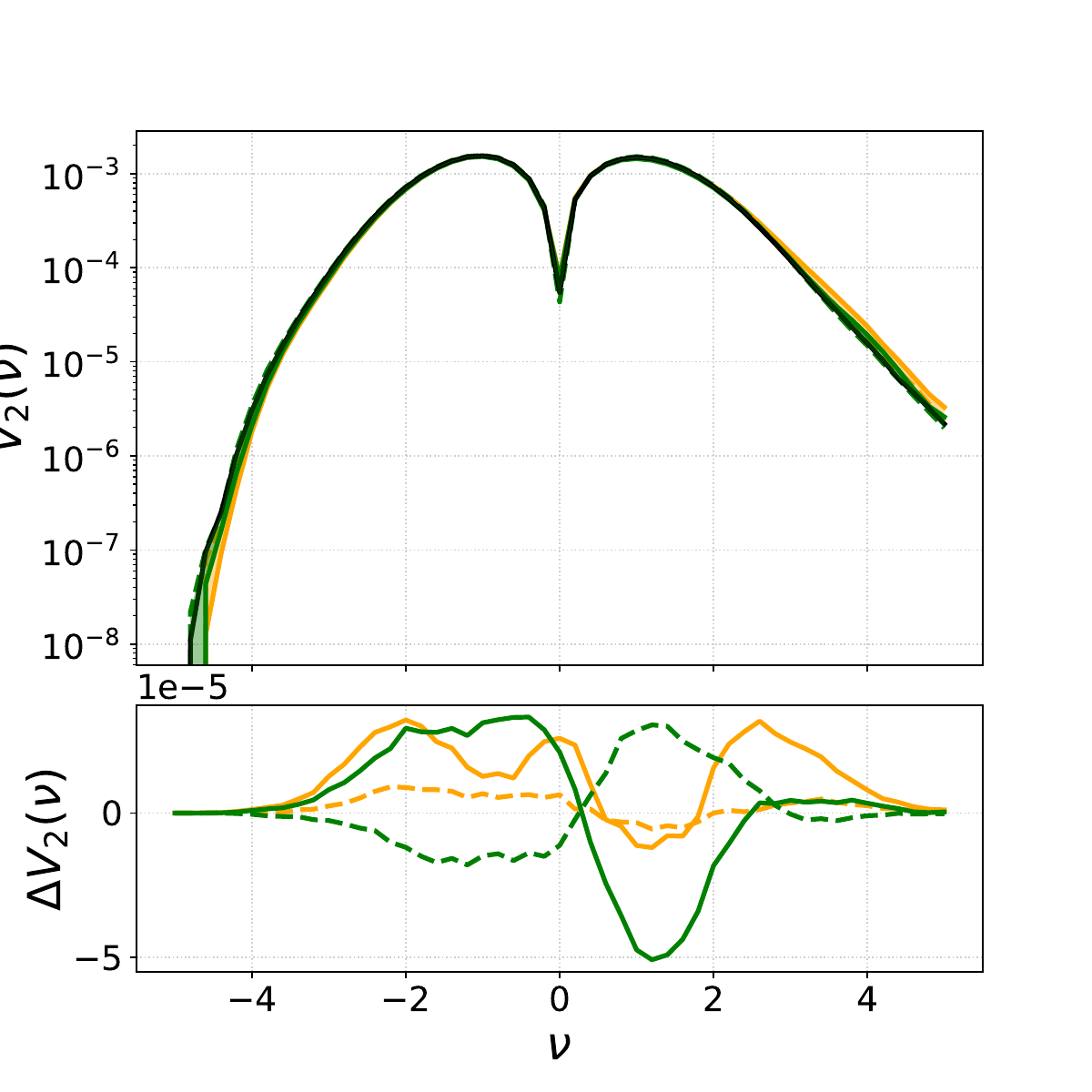}
 \caption{{\bf (A)} MFs for the LIM map for specific ranges of astrophysical parameters $a$ and $b$ for the ideal case without noise (top row) and with noise (bottom row), for the noise specification $\sigma_n=6000$ Jy sr$^{-1}$ and smoothing scale of 3 cMpc. The plots are average over 150 realizations. The fiducial case with $a=0.8475,b=7.22, A_I=0$ is shown in black. The smoothing scale is 3 cMpc. {\bf (B)} Same as {\bf (A)} but for  variations of parameters $A_I$ and $k_{\rm peak}$.  } 
    \label{fig:MF_vary_params}
\end{figure*}

As shown in the left panels of Fig. \ref{fig:histogram}, the [C\,{\sc ii}] LIM maps exhibit significant departures from Gaussianity. In particular, the intensity field is positive definite and exhibits a pronounced long positive tail, indicating a highly skewed distribution of intensity values. 
Adding Gaussian noise tends to Gaussianize the field, as seen in the right panels of Fig. \ref{fig:histogram}. Given the highly non-Gaussian and complex statistical nature of these fields, analytic expressions for the Minkowski functionals are not available. We therefore resort to numerical evaluation of the MFs directly from the intensity maps.

To compute the 2D MFs, we developed a numerical code based on the algorithm described in \citep{SG1998,GPJBBBS1990}. We validate the numerical accuracy of the code using 
an ensemble of Gaussian realizations and comparing the ensemble-averaged numerical MFs with the corresponding analytical predictions given by Eq. \ref{eq:MF_gaussian} with accuracy at the $\sim 1\%$ level.  

\subsection{Impact of model parameters on Minkowski functionals of LIM maps }
\label{sec:MF_B}

In this section, we focus exclusively on the case of $z=5.8$. We simulate ideal LIM maps and noise maps, which are then combined to generate signal-plus-noise maps, as described in Sec. \ref{sec:producing_maps}. All maps are smoothed at the appropriate scale.
Before evaluating the MFs, all maps are mean-subtracted and normalized by their respective standard deviations. The MFs are then computed over a normalized threshold range of $\nu\in [-5,5]$, with an increment of $\Delta\nu=0.2$. These calculations are performed at three smoothing scales: 1.6, 3, and 6 cMpc.

Fig.~\ref{fig:MF_std_with_noise} shows the MFs for the fiducial model ($a=0.8475,\, b=7.22,\, A_I=0$), comparing the ideal signal maps, the signal-plus-noise maps, and the pure noise maps.  By definition, $V_0$ is dimensionless, while $V_1$ and $V_2$ are given in units of cMpc$^{-1}$ and cMpc$^{-2}$, respectively. The error bars are the standard deviations obtained from 150 realizations. The smoothing scale used in this figure is 1.6 cMpc. Consistent with the observations discussed above, the amplitudes and shapes of the MFs for the signal-plus-noise maps closely resemble those of the pure Gaussian noise maps~\cite{Chingangbam:2023_composite}. This behavior reflects the low signal-to-noise ratio (SNR), which causes the morphological signatures of the underlying signal to be largely obscured by instrumental noise at this smoothing scale.

\subsubsection{Impact of astrophysical parameters $a,b$} 

From Eq.~\ref{eq:LCII2}, we infer that the parameter $a$ controls the power-law dependence on the SFR, whose spatial fluctuations determine those of the LIM field. We therefore expect $a$ to have a significant impact on both the shapes and amplitudes of the MFs. In contrast, $b$ enters only through a multiplicative constant factor in the LIM field. For a normalized Gaussian field, the analytic expressions for the MFs show that an overall multiplicative factor does not affect $V_0$, while the amplitudes of $V_1$ and $V_2$ depend on the ratio $\sigma_1/\sigma_0$, causing the multiplicative factor to cancel out. By analogy, we expect the MFs in the non-Gaussian case considered here to be substantially less sensitive to $b$, although they need not be completely independent of it.

The top row of Fig.~\ref{fig:MF_vary_params}(A) presents the MFs for the ideal LIM maps across selected ranges of the parameters $a$ and $b$, together with their deviations, $\Delta V_k$, relative to the fiducial model. These deviations are computed by varying one parameter at a time while keeping the remaining parameters fixed at their fiducial values. This approach allows us to isolate the individual effects of $a$ and $b$ on the morphological characteristics of the LIM maps. 
All maps are smoothed at a scale of 3 cMpc. A logarithmic scale is used for the $y$-axis to better highlight the differences between models. Accordingly, we plot $|V_2|$ instead of $V_2$ to accommodate its sign changes on the logarithmic scale. As expected, variations in $a$ produce substantial changes in both the shapes and amplitudes of the MFs, whereas the MFs exhibit considerably weaker sensitivity to variations in $b$.

The bottom row of Fig. \ref{fig:MF_vary_params}(A) shows the MFs and their relative deviations for the signal-plus-noise maps, using the same parameter ranges and smoothing scale as in the upper panels. Despite a low SNR, variations in $a$ produce significant deviations from the fiducial model. Although the amplitudes of these deviations are slightly suppressed, they remain comparable to those observed in the ideal case.  However, the inclusion of instrumental noise alters the shapes of the MFs, modifying the morphological signatures associated with variations in $a$. In contrast, the inclusion of noise substantially amplifies the sensitivity of the MFs to variations in $b$. \color{black} As we have stated earlier, the standard deviation of signal intensity maps is much smaller than standard deviation ($\sigma_n$) of instrumental noise. However, due to long tail of intensity distribution (Fig. \ref{fig:histogram}), the signal peeks out of noise at $\nu\gtrsim 2$ (Fig. \ref{fig:MF_vary_params}) for $\sigma_n=6000$ Jy sr$^{-1}$. Increasing $b$ makes it easier for the signal to stand out from noise of given $\sigma_n$. As we increase $\sigma_n$, we expect the signal to get buried inside the noise more and more and constraining power should decrease for both parameters $a$ and $b$. \color{black} 


At higher smoothing scales, the deviation amplitudes decrease for the ideal signal due to the suppression of small-scale structures, while the overall deviation shapes remain largely unchanged. For the signal-plus-noise case, increased smoothing further suppresses noise, causing the deviation shapes to approach those of the ideal signal.

These results indicate that the non-Gaussian information encoded in the MFs, particularly due to massive halos which produce luminous sources, is sensitive to the underlying astrophysical model even in the presence of substantial noise. Thus, MFs have the potential to detect and distinguish between different astrophysical model parameters.

\subsubsection{Impact of primordial parameters $A_I, k_{\rm peak}$} 

Next, we examine the impact of variations in the primordial feature parameters, $A_I$ and $k_{\rm peak}$. Since neither enters the field as a multiplicative factor, both are expected to alter the shape of the MFs. The top row of Fig. \ref{fig:MF_vary_params}(B) shows the MFs of the ideal LIM maps for selected parameter values and their deviations from the fiducial model. Each deviation is computed by varying one parameter at a time while keeping the others fixed at their fiducial values. The smoothing scale is again 3 cMpc. As expected, we find that both parameters significantly affect the amplitudes and shapes of the MFs.

The corresponding results for the signal-plus-noise maps are shown in the bottom row. Similar to the astrophysical parameters, the inclusion of noise amplifies the effects of both primordial parameters, increasing the deviation amplitudes and modifying their shapes.

Again, as expected, at higher smoothing scales the suppression of small-scale structures reduces deviation amplitudes in the ideal signal while preserving their shapes. For signal-plus-noise maps, additional noise suppression drives the deviation shapes towards those of the ideal signal.

Finally, a visual  comparison of the corresponding plots in (A) and (B) panels of Figs. \ref{fig:MF_vary_params} inform that the shape information of the MFs can potentially distinguish different astrophysical and primordial model parameters. This motivates us to quantify their distinguishability and carry out a Fisher matrix analysis in the next section.

\section{Projections for parameter constraints using Fisher matrix formalism}
\label{sec:fisher}

In this section, we estimate how well we can constrain the parameters of our model using MFs. We use the Fisher matrix formalism to do this computation. The Fisher matrix for each MF statistic $V_k$ can be written as,
\begin{equation}
    {\cal F}_{k,\mu\nu}=\left(\frac{\partial V_k}{\partial \theta_{\mu}}\right)(C_{k})^{-1}\left(\frac{\partial V_k}{\partial \theta_{\nu}}\right),
    \label{eq:fisher_matrix}
\end{equation}
where $\mu,\nu$ runs over the parameters of our model parameters, $C$ is the covariance matrix and $V_k$ are the MFs with $k=0,1,2$. The partial derivatives are numerically approximated by finite differences, as,
\begin{equation}
    \frac{\partial V_k}{\partial \theta_{\mu}}=\frac{1}{N_c}\sum_{n=1}^{N_c} \left(\frac{V_k(\theta_{\mu,2})-V_k(\theta_{\mu,1})}{\theta_{\mu,2}-\theta_{\mu,1}}\right)
    \label{eq:fisher_derivative}
\end{equation}
where $\theta_{\mu,1}$ is the fiducial value of parameters and $\theta_{\mu,2}$ is a slightly different choice of the same parameter, and $N_c$ is the number of simulations. 
The covariance matrix is expressed as,
\begin{equation}
    C_{k,ij}=\frac{1}{N_c-1}\sum_{n=1}^{N_c}\left[V_{k,n}(\nu_i)-\overline{V}_k(\nu_i)\right]\left[V_{k,n}(\nu_j)-\overline{V}_k(\nu_j)\right],
\end{equation}
where $i,j$ are index for $\nu$ grid and $\overline{V}_k$ is the average or mean $V_k$ over $N_c$ realizations.

 Note that non-zero, off-diagonal elements represent degeneracies between the model parameters. In order to obtain projected constraints on individual parameters, one has to marginalize over the other parameters. In general, this can be done as follows \citep{Matsubara2004}. Let us write the Fisher matrix as,  
 \begin{equation}
 {\cal F}= \begin{bmatrix}
  F & B\\ 
  B^T & D
\end{bmatrix},
\end{equation}
where $F$ is the submatrix of parameters that we are interested in, $D$ is the submatrix of parameters that we want to marginalize over, and $B$ are the cross-terms between the two sets of parameters. The marginalized or reduced Fisher matrix over the desired parameter set can then be obtained as, 
\begin{equation}
F^{\rm red}=F-BD^{-1}B^T.
\end{equation}
The projected marginalized 1$\sigma$ uncertainty for the parameter of interest is then given by $(F_{\mu}^{\rm red})^{-1/2}$. 

 To obtain well-converged  $\frac{\partial V_k}{\partial \theta_{\mu}}$  and  covariance  a sufficiently large number of simulations $N_c$ is required, with the rule of thumb being $N_c \gg p$, where $p$ is the data vector size. In our context, $p$ is in general given by the product of the number of MFs (if we cross-correlate across different MFs), number of thresholds, number of smoothing scales, and so on.

Generating a large number of simulations is computationally expensive. Therefore, we compute $C_k$ independently at each smoothing scale, without evaluating cross-correlations between different smoothing scales. The normalized covariance matrix is defined as 
$r_{k,ij}=\frac{C_{k,ij}}{\sqrt{C_{k,ii}C_{k,jj}}}$. 
We find that the correlations between the threshold bins decrease progressively from $V_0$ to $V_2$, with the normalized covariance matrix for $V_2$ becoming substantially more diagonal than those for $V_0$ and $V_1$, consistent with the findings of \cite{Bashir:2026}. Motivated by this reduced correlation, we perform the Fisher analysis using only $V_2$. 
Furthermore, since the true LIM signal is positive definite, we expect the signal at negative thresholds, $\nu\lesssim 0$, to be dominated by noise. We therefore restrict the Fisher analysis to the threshold range $0.2\leq\nu\leq5$, corresponding to $p=25$ threshold points.

\begin{figure}[t]
        \centering
        \includegraphics[width=.9\columnwidth]{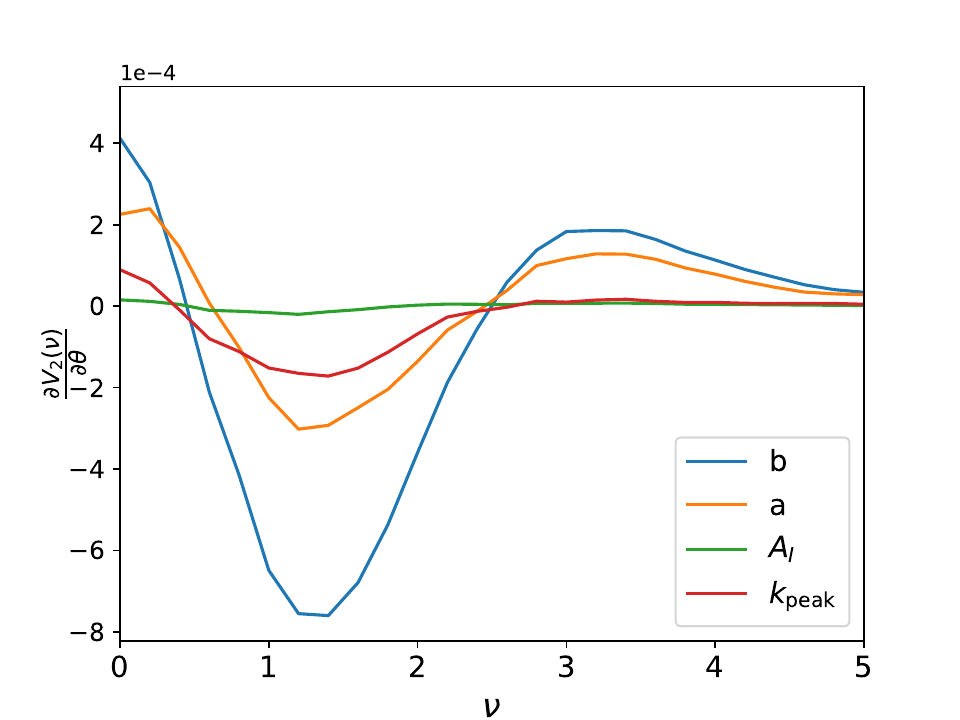}
        \caption{$\frac{\partial V_2}{\partial\theta_{\mu}}$ where $\theta_{\mu}$ is the four parameter model including the modification to inflationary power spectrum. The parameter $A_I$ is chosen in units of $10^{-9}$. For this calculation, we have chosen $\sigma_n=6000$ Jy sr$^{-1}$ with smoothing scale= 3 cMpc.}
        \label{fig:MF_vs_params}
\end{figure}

To compute $\frac{\partial V_k}{\partial\theta_{\mu}}$, we use $N_c=20$, since generating a large number of simulations for different values of the model parameters is computationally extremely expensive. The resulting derivatives, $\frac{\partial V_2}{\partial\theta_{\mu}}$, for each of the four parameters, $a$, $b$, $A_I$, and $k_{\rm peak}$, are shown in Fig. \ref{fig:MF_vs_params}. We find that the derivatives are relatively stable, with limited statistical fluctuations across the threshold bins, indicating that $N_c=20$ provides sufficiently stable estimates of the derivatives.

Finally, to fix $N_c$ for the covariance matrix $C_{k=2}$, we compute the covariance for different values of $N_c$ and examine the convergence of the parameter uncertainties obtained from the Fisher matrix as $N_c$ is increased. The results of this convergence test are shown in Fig. \ref{fig:convergence_test} in Appendix \ref{sec:cov}. We find that the parameter uncertainties are well converged for $N_c=150$. We therefore adopt $N_c=150$ for the covariance matrix and use this value throughout our analysis.

In what follows, we consider two physical scenarios: (i) our fiducial astrophysical model, in which only the two astrophysical parameters $a$ and $b$ are varied, and (ii) an extended model in which, in addition to the astrophysical parameters, we allow for non-zero values of $A_I$ and $k_{\rm peak}$. 
For both scenarios, we carry out the Fisher analysis by including instrumental noise to the simulated signals. We consider noise configurations of CCAT  with $\sigma_n=6000$ Jy sr$^{-1}$ at $z=5.8$. Additionally, we repeat the analysis at $z=7.6$ with $\sigma_n=4000$ Jy sr$^{-1}$. 

\subsection{Constraints on the astrophysical model}
\label{sec:astro_model}

We now consider the first scenario where we constrain the astrophysical parameters $a$ and $b$, with the primordial power spectrum given by the standard featureless power law form. The input fiducial model has parameter values $(b,a)=(b_0,a_0)=(7.22,0.8475)$. Elements of the Fisher matrix ${\cal F}$, and the corresponding reduced matrix $F^{\rm red}$,  which in this case is a $2\times 2$ matrix, computed using Eq. \ref{eq:fisher_matrix} are shown in Table \ref{tab:table_1}. The index 0 corresponds to $b$, while 1 corresponds to $a$. This table serves to demonstrate the negative correlation between $a$ and $b$ (it is easy to check that ${\cal F}^{-1}$ has negative off-diagonal terms), and its impact on the reduced Fisher matrix elements which are significantly reduced relative to the the elements of ${\cal F}$. 

\begin{table}[ht]
\caption{The Fisher matrix ${\cal F}$, and the corresponding reduced matrix $F^{\rm red}$, for the two parameter astrophysical model, computed for for $V_2$ in the threshold range $0\leq \nu\leq 5$. The index 0 corresponds to parameter $b$, while 1 corresponds to $a$. The smoothing scale is denoted by $\ell_s$.} 
\label{tab:table_1}
    \centering
    \renewcommand{\arraystretch}{1.5}
       \begin{tabular}{|c|c|c|c|}
        \hline
        \multirow{2}{*}{ ${\cal F}$ and $F^{\rm red}$
        }
        & \multicolumn{3}{c|}{$z=5.8,\ \sigma_n=6000$ Jy sr$^{-1}$} \\ 
        \cline{2-4}
         &\  $\ell_s$ = 1.6 cMpc \ & \  $\ell_s$ = 3 cMpc \ & \ $\ell_s$ = 6 cMpc \ \\
        \hline
        ${\cal F}_{00}$ & 262.15 & 259.41 & 180.89 \\
        ${\cal F}_{01} = {\cal F}_{10}$ & 105.67 & 75.48  & 69.39  \\
            ${\cal F}_{11}$ & 64.57  & 62.54  & 47.62  \\ \hline\hline
                $F_{0}^{\rm red}$ & 89.22 & 168.3 & 79.78 \\
    $F_{1}^{\rm red}$ & 21.98 & 40.57 & 21.00 \\ 
    \hline
    \end{tabular}
   \end{table}

\begin{figure}[t]
        \centering
        \includegraphics[width=0.85\linewidth]{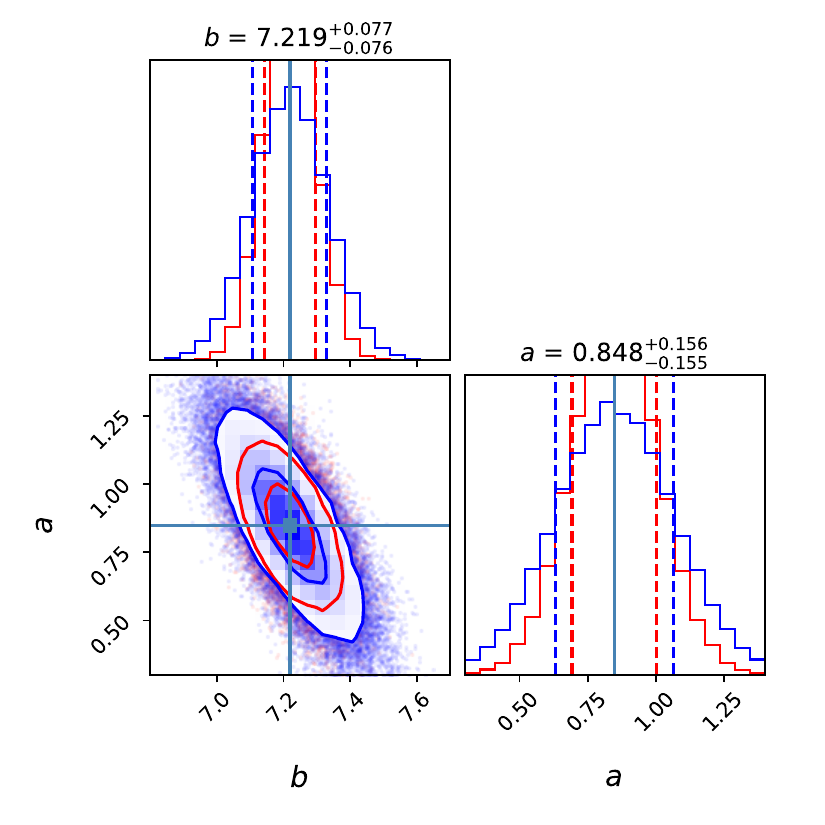}\\
        \caption{Projected constraints (at 1$\sigma$ level) on fiducial astrophysical parameters $a$ and $b$ using the Fisher matrix components in Table \ref{tab:table_1} with the scenario $\sigma_n=6000$ Jy sr$^{-1}$ and smoothing scale of 3 cMpc (red) and 6 cMpc (blue). The 1$\sigma$ error bars quoted in the figure corresponds to smoothing scale 3 cMpc case. The vertical red and blue dashed lines mark the mean and 1$\sigma$ limits.} 
        \label{fig:fisher_std}
\end{figure}

In Fig. \ref{fig:fisher_std}, we plot the projected 2D probability contours obtained by sampling $a$ and $b$ as mutivariate Gaussians with means $a_0, b_0$, and covariance matrix $\Sigma={\cal F}^{-1}$. The 1$\sigma$ and 2$\sigma$ contours for smoothing scale 3 cMpc are shown by the red curves.   For comparison, we also show the corresponding contours for the case of smoothing 6 cMpc (blue curves). The corresponding marginalized 1D constraints at 1$\sigma$  level are also shown. As mentioned above, we observe that $a$ and $b$ are anti-correlated, with the direction of correlation unaffected by the change of smoothing scale.   


The constraints on $b$ and $a$ are $7.22\pm 0.08$ and $0.84\pm 0.16$, respectively, at 1$\sigma$ confidence level for smoothing scale of 3 cMpc.The recovered mean values are quite close to the input fiducial values, implying that the addition of Gaussian noise does not bias their inferred central values. This is an encouraging result implying that with the noise cofiguration of {\it CCAT} observatory there is a good prospect of constraining parameterized astrophysical models. Although the SNR is low, the constraints are driven by $\nu>0$ where the approximately lognormal high tails of the signal field dominate the Gaussian noise for which high $\nu$ values are exponentially suppressed.

An interesting feature of the $F^{\rm red}$ values in Table \ref{tab:table_1} is that increasing the smoothing scale from 1.6 to 3 cMpc tightens the constraints on both $a$ and $b$. This can be attributed to the more effective suppression of noise at the higher smoothing scale, while preserving the underlying small-scale structures of the ideal signal. In contrast, increasing the smoothing scale further from 3 to 6 cMpc degrades the constraints, as the small-scale structures in the LIM maps are progressively washed out. Thus, a smoothing scale of 3 cMpc emerges as the optimal choice, minimizing the uncertainties by balancing noise suppression against information loss due to the erasure of small-scale structures. This finding highlights the importance of performing the analysis at multiple smoothing scales when working with observational data.

\begin{figure}[t]
        \centering
        \includegraphics[width=0.85\linewidth]{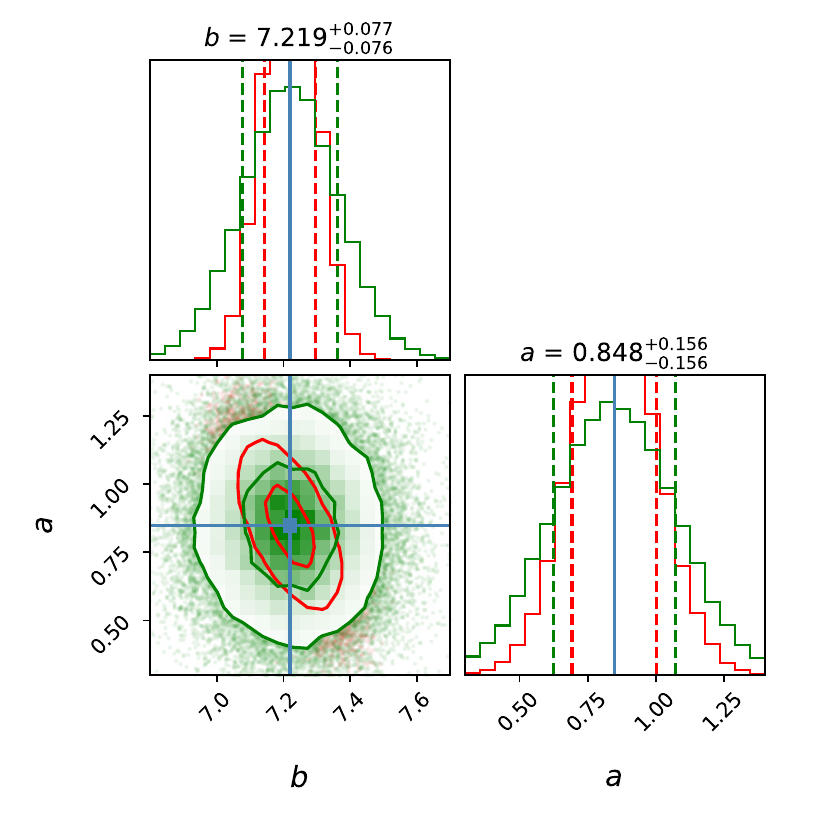}
        \caption{Same as Fig. \ref{fig:fisher_std} but for two $z$ at 5.8 (red) and 7.6 (green) respectively. $\sigma_n=6000$ and $4000$ Jy sr$^{-1}$ for $z$=5.8 and 7.6, respectively. We have used a smoothing scale of 3 cMpc. The 1$\sigma$ error bars quoted in the figure corresponds to $z=5.8$ case. The vertical red and green dashed lines mark the mean and 1$\sigma$ limits. } 
        \label{fig:fisher_std_z}
\end{figure}

Next, we repeat the analysis at $z=7.6$ and compare the results with those obtained at $z=5.8$, using a smoothing scale of 3 cMpc in both cases. The projected 2D and 1D probability contours are shown in Fig. \ref{fig:fisher_std_z}. We find that the constraints are generally slightly weaker at the higher redshift, despite the lower noise standard deviation. This is expected because the line intensity decreases at higher redshifts owing to the reduced abundance of host dark matter halos, leading to a lower signal-to-noise ratio (SNR). However, an important feature is that the orientation of the degeneracy between $a$ and $b$ differs between the two redshifts. This suggests that combining information from multiple redshifts could help break parameter degeneracies and thereby yield tighter constraints on the model parameters.

\subsection{Constraints on the extended astrophysical-plus-primordial feature model}
\label{sec:extended_model}
\begin{figure}[]
        \centering
        \includegraphics[width=\linewidth]{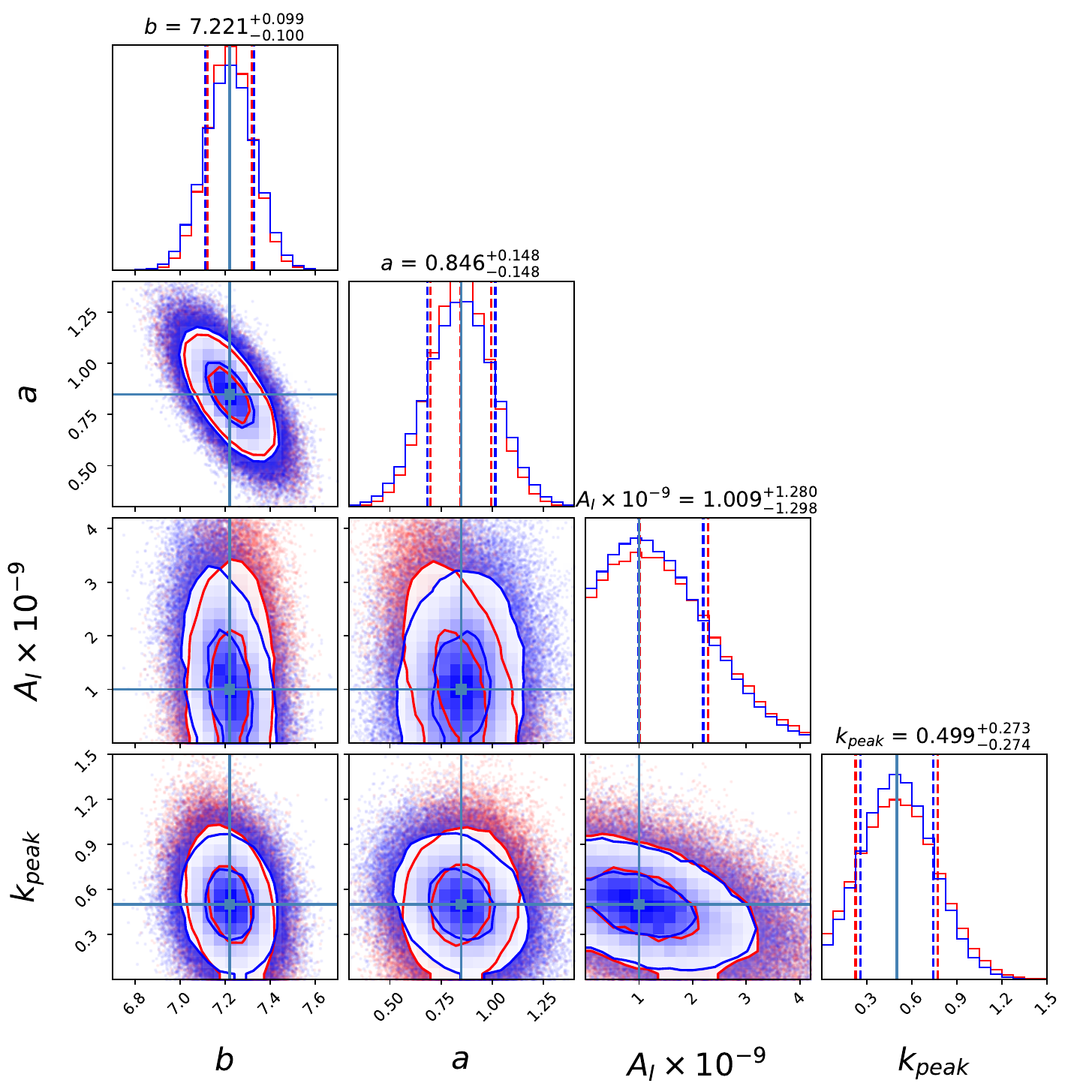}
                \caption{Projected 1$\sigma$ constraints on fiducial astrophysical parameters $a$ and $b$, as well as modifield inflationary parameters $A_I$ and $k_{\rm peak}$ in the scenario $\sigma_n=6000$ Jy sr$^{-1}$ and smoothing scale of 3 cMpc (red) and 6 cMpc (blue). The 1$\sigma$ error bars quoted in the figure correspond to smoothing scale 3 cMpc case.  The vertical red and blue dashed lines mark the mean and 1$\sigma$ limits.} 
        \label{fig:fisher_with_AI}
\end{figure}

\begin{figure}[]
       \centering
        \includegraphics[width=\columnwidth]{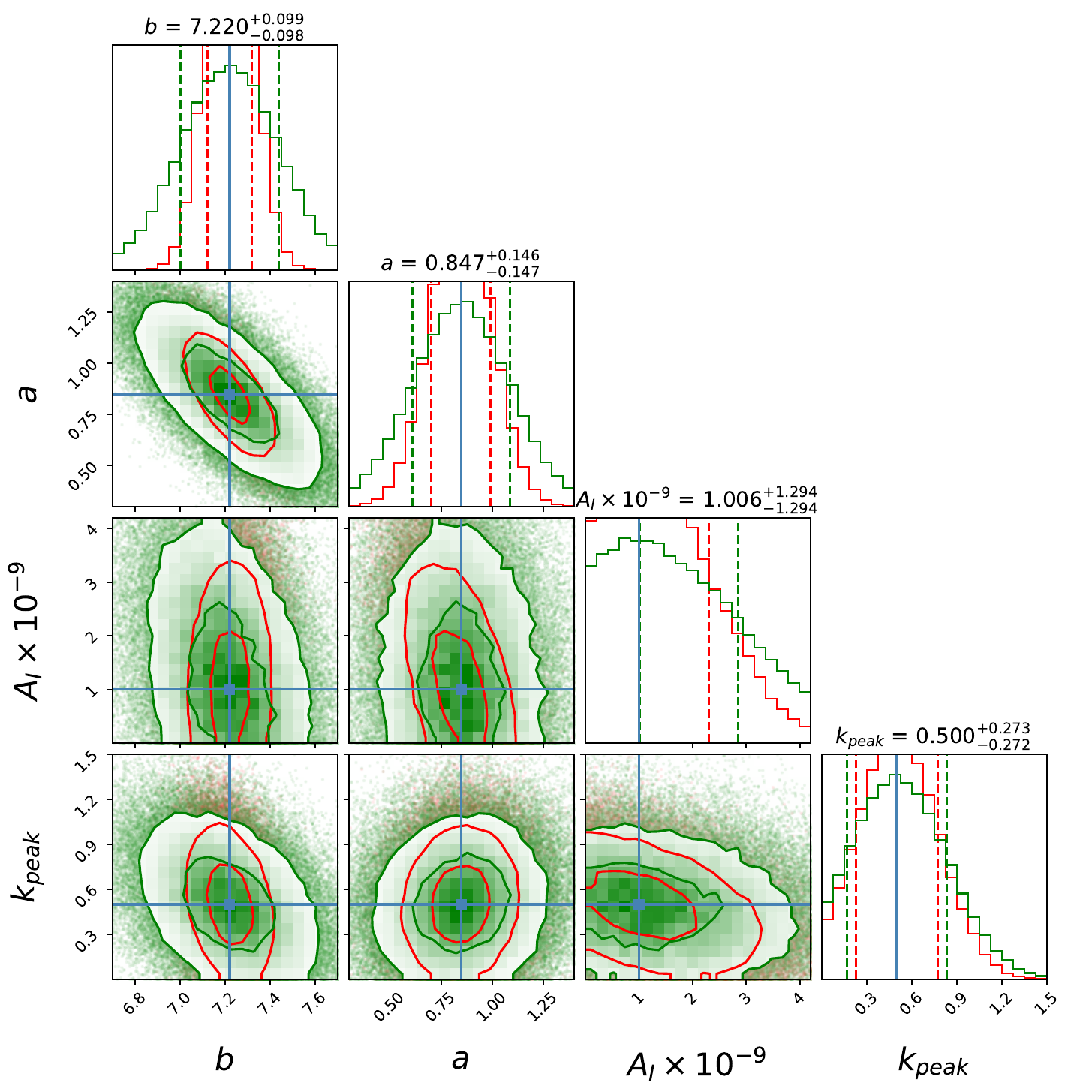}
        \caption{Same as Fig.~\ref{fig:fisher_with_AI}, but showing results for $z=5.8$ (red) and $z=7.6$ (green). The instrumental noise levels are $\sigma_n=6000$ and $4000\,\mathrm{Jy\,sr^{-1}}$ at $z=5.8$ and $z=7.6$, respectively, and a smoothing scale of $3\,\mathrm{cMpc}$. The $1\sigma$ error bars correspond to the $z=5.8$ case. The vertical red and green dashed lines indicate the mean values and their respective $1\sigma$ limits.}
        \label{fig:fisher_with_AI_z}
\end{figure}

We next consider a second scenario in which we simultaneously constrain the astrophysical parameters $a$ and $b$, along with the primordial feature parameters $A_I$ and $k_{\rm peak}$. We adopt the fiducial parameter values $(b,a,A_I,k_{\rm peak})=(7.22,0.8475,10^{-9},0.5\,{\rm cMpc})$. Our choice of $A_I$ is motivated by current constraints from CMB and LSS observations, which place upper limits of $A_I\lesssim 10^{-10}-10^{-9}$ (95\% CL) on this particular model at $k\lesssim 0.1\,{\rm Mpc}^{-1}$ \citep{NFC2022,BFMMV2023}. It is therefore interesting to investigate whether signatures of a non-zero $A_I$ can be detected at larger wavenumbers. Recent studies have demonstrated that the 21 cm signal from the epoch of reionization can potentially probe non-zero $A_I$ at these smaller scales \citep{NCSMF2025,VNC2026}. The 21 cm signal probes the properties of the intergalactic medium (IGM) during reionization, including its temperature and ionization fraction, as well as the astrophysical properties of galaxies, such as their Lyman-$\alpha$ and X-ray emissivities. In contrast, the intensity of [C\,II] line emission is less sensitive to IGM properties and may therefore provide a more direct probe of such non-standard primordial scenarios.

In Fig. \ref{fig:fisher_with_AI}, we plot the projected 2D probability contours obtained by sampling $a$, $b$, $A_I$ and $k_{\rm peak}$ as mutivariate Gaussians with means given by their fiducial values quoted above. The covariance matrix $\Sigma={\cal F}^{-1}$ in this case is a $4\times 4$ matrix. The 1$\sigma$ and 2$\sigma$ contours for smoothing scale 3 cMpc are shown by the red curves, while the corresponding contours for smoothing scale 6 cMpc is shown by the blue curves. The corresponding marginalized 1D constraints at 1$\sigma$  level are also shown.

Similar to the case in which only the astrophysical parameters are constrained, we find that $a$ and $b$ are anti-correlated, with the direction of degeneracy remaining unchanged across smoothing scales. Visual inspection also reveals that the degeneracies between the astrophysical and primordial feature parameters depend on the smoothing scale. Specifically, $A_I$ is largely uncorrelated with $b$ at a smoothing scale of 3 cMpc, but exhibits a mild anti-correlation at 6 cMpc. The trend is reversed for the degeneracy between $A_I$ and $a$, which changes from mild anti-correlation to mild correlation as the smoothing scale increases from 3 to 6 cMpc. In contrast, $k_{\rm peak}$ exhibits a mild anti-correlation with $b$ at both smoothing scales, while its degeneracy with $a$ changes from mild correlation at 3 cMpc to mild anti-correlation at 6 cMpc.

The constraints on $b, a, A_I\times 10^{-9}$ and $k_{\rm peak}$ are $7.221\pm 0.100$, $0.846\pm 0.148$, $1.009 \pm 1.298$ and $0.499\pm 0.274$,  respectively, at 1$\sigma$ CL for smoothing scale of 3 cMpc. All the recovered mean values are close to the input fiducial values, again implying that the addition of Gaussian noise does not introduce bias in  their inferred central values. The $1\sigma$ uncertainty for $b$ is about $25\%$  higher, while that of $a$ is lower by about 5\%, compared to when only the astrophysical parameters are constrained.  Despite a well-defined central value, $A_I$ is unconstrained on the lower side. In comparison, $k_{\rm peak}$ is well constrained at $1\sigma$  with the uncertainty being about 50\% of the central value.  The uncertainties for all four parameters are largely unchanged for the higher smoothing scale of 6 cMpc.

Comparing the analysis at $z=7.6$ with that at $z=5.8$, for a fixed smoothing scale of 3 cMpc, we find that the uncertainties for all four parameters are higher with no bias of the inferred central values. We again observe small but discernible changes of the orientations of some parameter degeneracies with respect to redshift which may be exploited to minimize degeneracies. 

From the results of this subsection, we conclude that the noise characteristics of the {\it CCAT} observatory offer the potential to constrain the astrophysical and primordial parameters simultaneously, despite the low SNR. Incorporating observational data across multiple smoothing scales and redshifts can further improve the constraints. As in the previous case, the constraints are primarily driven by the $\nu>0$ regime, owing to the primordial parameters have a pronounced impact in this threshold range due to the enhanced abundance of high-mass halos, in addition to the approximately lognormal high-$\nu$ tails of the signal field dominate over the Gaussian noise.

We emphasize that, at the observing frequencies of 220 and 280 GHz adopted in this work, the [C\,{\sc ii}] signal may be contaminated by foreground emissions, particularly Galactic thermal dust emission, as well as emission from extragalactic point sources, such as dusty galaxies. In our analysis, we assume perfect foreground subtraction and do not account for residual contamination. Any residual foreground emission present may bias our results and should therefore be considered when interpreting our findings.

\section{Conclusions}
\label{sec:conclusions}

Line intensity mapping probes the cumulative emission from galaxies across the Universe in specific spectral lines, with the observed frequency encoding the redshift of the emitting sources. Since galaxies form within dark matter halos, whose spatial distribution is inherently non-Gaussian, the resulting intensity maps are also expected to exhibit significant non-Gaussian features. In this work, we employ Minkowski functionals to characterize the non-Gaussianity of line intensity maps. 

We use the semi-numerical code \texttt{21cmFAST}, together with the experimental configuration of \texttt{CCAT}, to generate [C\,{\sc ii}] intensity maps at $z=5.8$ and $z=7.6$. We adopt a simple two-parameter astrophysical model that relates the [C\,{\sc ii}] line luminosity to the mass of its host dark matter halo, along with a prescription for star formation. We find that the Minkowski functionals $V_0$, $V_1$, and $V_2$ exhibit pronounced deviations from Gaussian predictions at high thresholds ($\nu$). Remarkably, despite instrumental noise, these statistics retain clear signatures of non-Gaussianity, deviating significantly from the expectations for a purely Gaussian noise field at relatively high positive thresholds. Using the $V_2$ statistic within a simple Fisher matrix formalism, we demonstrate that the upcoming \texttt{CCAT} experiment can place meaningful constraints on the underlying astrophysical parameters.

We further investigate the potential of [C\,{\sc ii}] intensity mapping to detect a bump-like feature in the primordial power spectrum generated by particle production during inflation. This extended model introduces two additional parameters, $A_I$ and $k_{\rm peak}$, to characterize the feature's amplitude and characteristic wavenumber. Current constraints on such inflationary models are primarily derived from observations of CMB anisotropies and large-scale structure at wavenumbers $k\lesssim 0.1\,\mathrm{Mpc}^{-1}$, with sensitivity to feature amplitudes of $A_I\sim 10^{-10}$-$10^{-9}$. Our results indicate that [C\,{\sc ii}] line intensity mapping can extend these constraints to substantially smaller scales (higher wavenumbers), reaching amplitudes of $A_I\sim 10^{-9}$. This complements a recent investigation of the same class of inflationary models using 21-cm fluctuations during the EoR~\citep{VNC2026}.

For both the astrophysical model and the extended model incorporating a primordial feature, we find that combining measurements across multiple smoothing scales and redshifts can further tighten parameter constraints. This improvement arises from changes in the directions of parameter degeneracies across scales and redshifts, together with the trade-off between noise suppression and the loss of small-scale information as the smoothing scale increases. Furthermore, the long-tailed distribution of intensity fluctuations allows the non-Gaussian signal to dominate over instrumental noise in the high-threshold regime, even when the overall signal-to-noise ratio (SNR) is low. These results highlight the potential of Minkowski functionals to extract astrophysical and primordial information from [C\,{\sc ii}] intensity maps and to constrain model parameters with upcoming observations.

Our analysis relies on several assumptions that may affect the projected constraints quantitatively. The constraints are primarily driven by the approximately lognormal nature of the intensity field, which exhibits a long tail at high thresholds, whereas the contributions of Gaussian instrumental noise to the Minkowski functionals $V_k$ decay exponentially. Deviations from Gaussianity in the noise statistics could therefore alter the resulting constraints, and their impact warrants further investigation. Another limitation of our analysis is the neglect of residual foreground contamination in the LIM maps, effectively assuming that foreground removal is sufficiently efficient to isolate the cosmological signal. This assumption is idealized, and future studies should incorporate residual foreground contamination to assess its impact on the detectability of the signal and the resulting parameter constraints. Furthermore, combining Minkowski functionals with conventional statistical measures, such as the power spectrum and bispectrum, could enable a more complete extraction of cosmological information from the data. Finally, cross-correlating independently observed datasets, such as LIM and 21-cm intensity maps, could enhance the prospects for constraining astrophysical models, detecting primordial feature signals and provide tighter constraints in the event of a non-detection.

\begin{acknowledgments}
We acknowledge the use of the publicly available software packages \texttt{21cmFAST}, \texttt{emcee}, and \texttt{HMF}, as well as publicly available data from the Hubble Space Telescope (HST).
\end{acknowledgments}

\appendix

\section{Data and analysis of UVLF data}
\label{sec:data}
\begin{figure}[h]
\centering
\includegraphics[scale=0.24]{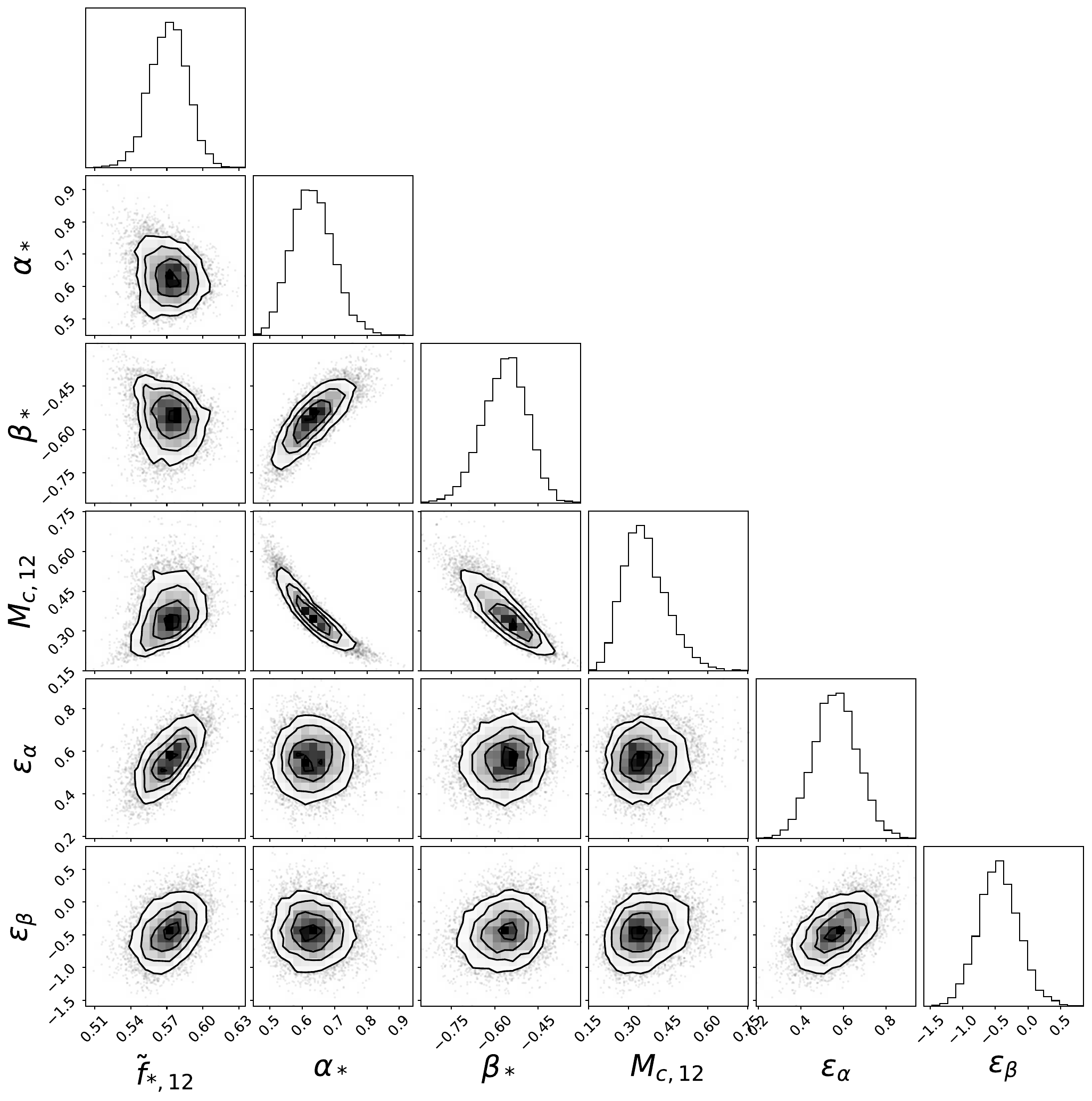}
\caption{Corner plot for our 6 parameter star formation model in Sec. \ref{sec:UVLF_modelling}. The results are reproduced from \citep{Acharya2026}. } 
\label{fig:corner_plot}
\end{figure}

To fit our astrophysical model in Sec. \ref{sec:UVLF_modelling} , we use the UVLF measurements made with Hubble Space Telescope (HST) within the redshift range 4 to 8. We note that our required redshift of $z=6$ for LIM analysis falls within this regime. The analysis follows \citep{Acharya2026} and briefly explained here for completeness. The measurements are in the magnitude regime $-23\lesssim M_{UV}\lesssim -16$. We have excluded few data points with upper limits, for the analysis. 

We use Gaussian likelihood to do our parameter estimation analysis. The log-likelihood is given as,
\begin{equation}
    -2{\rm ln}\mathcal{L}=\sum_{M_{UV},z}\left(\frac{\Phi_{UV}^{\rm data}-\Phi_{UV}^{\rm theory}}{\sigma_{\Phi_{UV}}}\right)^2
    \label{eq:log_likelihood}
\end{equation}
where $\Phi_{UV}^{\rm theory}$ is obtained from Eq. \ref{eq:Phi_theory}. We sum over contribution from individual data points over $M_{UV}$ and redshifts bins. We use the publicly available {\it emcee} sampler \citep{emcee2013} to run Markov Chain Monte Carlo (MCMC) using the Gaussian likelihood described above. Our prior choices and  posteriors are shown in Table \ref{tab:parameter_table}. We choose flat prior on all the parameters within the denoted range.

\begin{figure}
\centering
\includegraphics[scale=0.5]{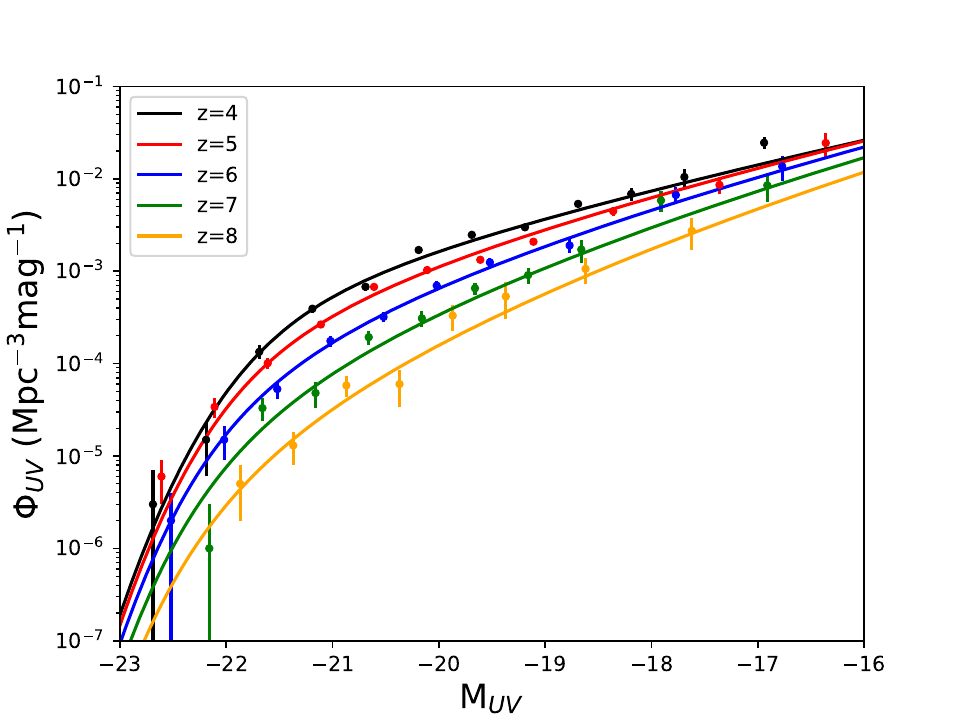}
\caption{Fit of our astrophysical model to the UVLF data using the best fit posterior values in Table \ref{tab:parameter_table}. The results are reproduced from \citep{Acharya2026}.  }
\label{fig:fiducial_fit}
\end{figure}

\vskip .3cm
\section{Convergence of parameter uncertainties}
\label{sec:cov}

\begin{figure}[h] 
\centering
 \includegraphics[width=.47\linewidth,height=1.35in]{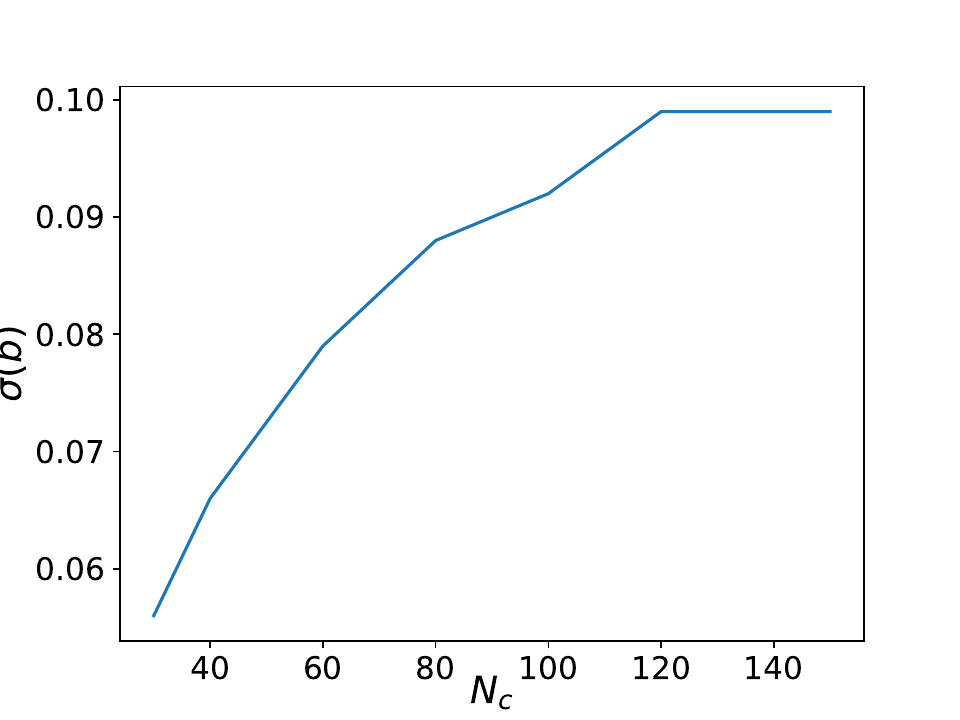}
 \includegraphics[width=.47\linewidth,height=1.35in]{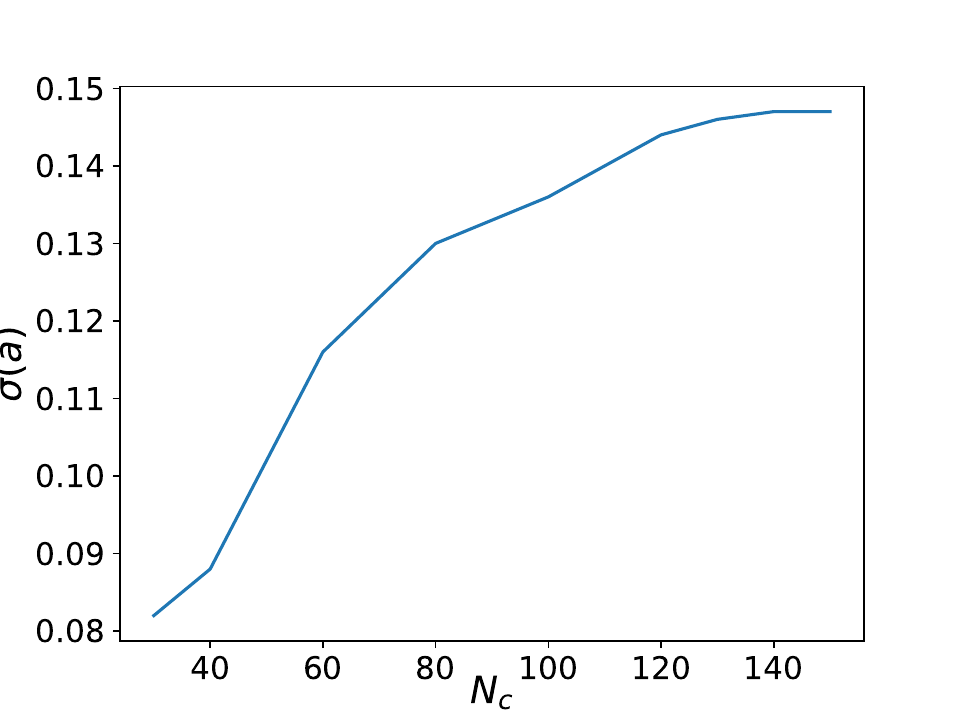} \\\includegraphics[width=.47\linewidth,height=1.35in]{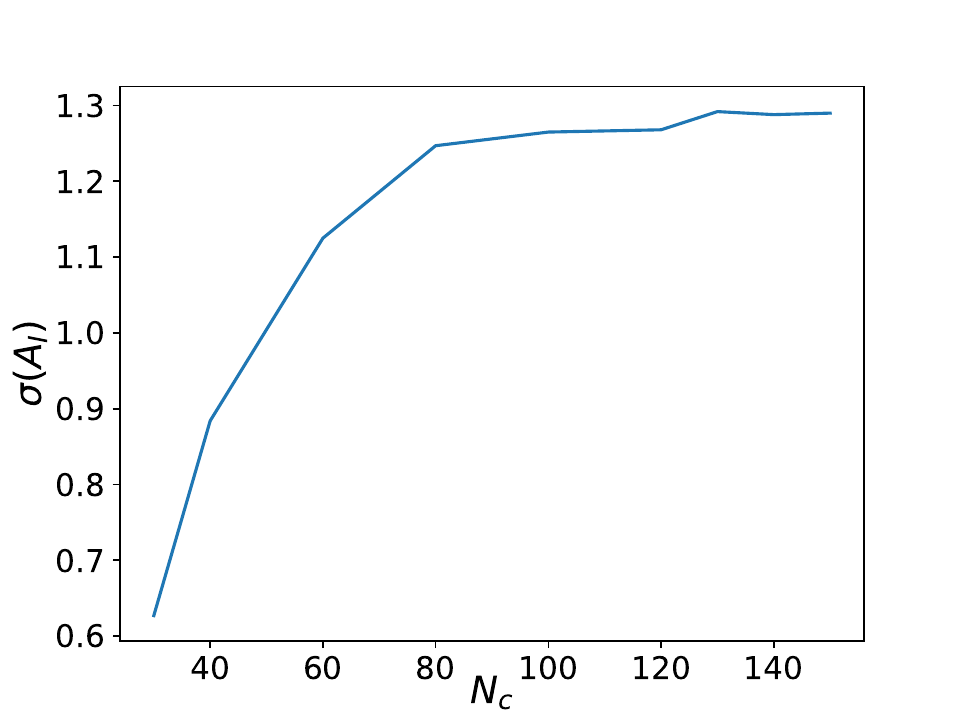}
  \includegraphics[width=.47\linewidth,height=1.35in]{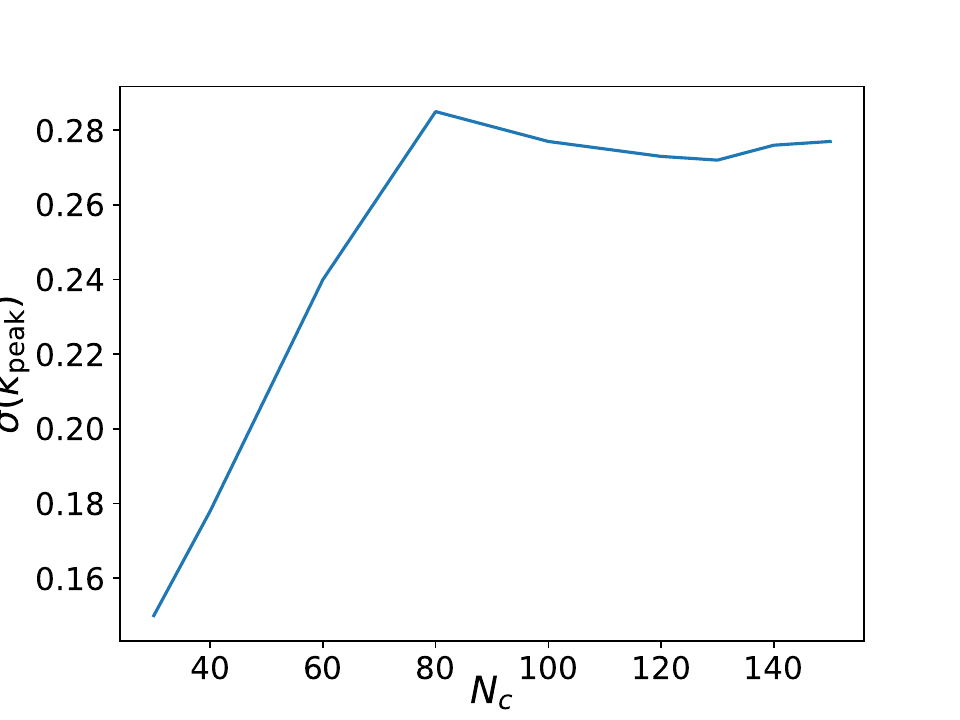}
  \caption{1$\sigma$ marginalized uncertainties for for each of the astrophysical and primordial parameter as a function of the number of simulations $N_c$, for $z=5.8$ and smoothing scale of 3 cMpc.}
  \label{fig:convergence_test}
\end{figure}

The parameter uncertainties obtained from a Fisher matrix analysis depend sensitively on the number of simulations, $N_c$, used to estimate the covariance matrix of the statistic. As discussed in Sec. \ref{sec:fisher}, these uncertainties typically approach convergence as $N_c$ increases sufficiently.

Here, we test the convergence of the parameter uncertainties for the extended astrophysical-plus-primordial feature model (see Sec. \ref{sec:extended_model}) by progressively increasing $N_c$. Figure \ref{fig:convergence_test} shows that the uncertainties are reasonably well converged for all parameters for $N_c \sim 150$. We therefore adopt this value for the results presented in Sec. \ref{sec:fisher}.

%


\bibliography{main}

\end{document}